\documentclass[jgrga]{AGUTeX}
\usepackage{amsmath}
\usepackage{amsfonts}
\usepackage{amssymb}
\usepackage{graphicx}
\usepackage{verbatim}

\authorrunninghead{DIMANT AND OPPENHEIM}

\titlerunninghead{M-I TURBULENT COUPLING: ENERGY BUDGET}

\authoraddr{Y. S. Dimant,
Center for Space Physics, Boston University, 725 Commonwealth Ave., Boston, MA 02215, USA.
(dimant@bu.edu)}

\begin{document}

\title{Magnetosphere-Ionosphere Coupling Through \emph{E}-region Turbulence: Energy Budget}

\author{Y. S. Dimant and M. M. Oppenheim}

\begin{abstract}

During periods of intense geomagnetic activity, strong electric
fields and currents penetrate from the magnetosphere into
high-latitude ionosphere where they dissipate energy, form
electrojets, and excite plasma instabilities in the
\emph{E}-region ionosphere. These instabilities give rise to
plasma turbulence which induces non-linear currents and strong
anomalous electron heating (AEH) as observed by radars. These
two effects can increase the global ionospheric conductances.
This paper analyzes the energy budget in the electrojet, while
the companion paper applies this analysis to develop a model of
anomalous conductivity and frictional heating useful in
large-scale simulations and models of the geospace environment.
Employing first principles, this paper proves for the general
case an earlier conjecture that the source of energy for plasma
turbulence and anomalous heating equals the work by external
field on the non-linear current. Using a two-fluid model of an
arbitrarily magnetized plasma and the quasilinear
approximation, this paper describes the energy conversion
process, calculates the partial sources of anomalous heating,
and reconciles the apparent contradiction between the
inherently 2-D non-linear current and the 3-D nature of AEH.

\end{abstract}

\begin{article}

\section{Introduction}

At high latitudes, the large-scale electric field,
$\vec{E}_{0}$, from the solar wind and Earth's magnetosphere
penetrates to the ionosphere. Across the
\emph{E}/upper\emph{D}-region altitudes, roughly between 80 and
130 km, the electrons are strongly magnetized, while the ions
become at least partially demagnetized due to frequent
collisions with the neutral atmosphere. This demagnetization
slows the drifting ions down, making the
$\vec{E}_{0}\times\vec{B}_{0}$ drift of electrons into strong
Hall currents named high-latitude electrojets. In the global
picture of magnetosphere-ionosphere (MI) coupling, this is the
region where field-aligned magnetospheric currents close and
dissipate energy.

Electrojet currents often drive plasma instabilities that
generate electrostatic field fluctuations coupled to plasma
density irregularities. These fluctuations have low
frequencies, usually smaller than the average frequencies of
electron and ion collisions with neutrals, $\nu_{e,i}$, while
the corresponding wavelengths in the unstable range exceed the
ion mean free path. Density irregularities, usually in the
range from tens of centimeters to tens of meters, are routinely
detected as strong coherent radar echoes
\citep[e.g.,][]{Cohen:Secondary67,Balsley:Radar71,Crochet:HF79,Kudeki:Condor87}.
Rocket flights through the lower ionosphere have detected also
electrostatic field fluctuations
\citep[e.g.,][]{Pfaff:Electric87b,Pfaff:ERRRIS,Pfaff:97a,Rose:92,Fukao:SEEK}.
These irregularities and fluctuations are caused by variety of
instabilities including the Farley-Buneman (FB)
\citep{Farley:Ejet63,Buneman:Ejet63}, gradient drift
\citep{Hoh:63,Maeda:Theoretical63}, and thermal instabilities
\citep{Dimant:95c,Dimant:Physical97,Kagan:thermal00,DimOppen2004:ionthermal1}.
The FB instability is excited when the relative velocity
between the average electron and ion streams exceeds the local
ion-acoustic speed. At high latitudes, this usually occurs when
$E_{0}\equiv |\vec{E}_{0}|$ exceeds the threshold value of
about $20$~mV/m. These and much stronger fields are not
uncommon in the sub-auroral, auroral, and polar cap areas,
especially during magnetospheric storms and substorms.

Small-scale fluctuations generated by these instabilities can
cause enormous anomalous electron heating (AEH). For about
thirty years, radars have observed strong electron temperature
elevations from 300-500~K up to more than 4000~K, correlating
with strong convection fields $\vec{E}_{0}$
\citep{Schlegel:81a,Providakes:88,Stauning:89,St.-Maurice:90a,Williams:1992,
Foster:Simultaneous00,Bahcivan:Plasma2007}. Simple estimates
show that regular ohmic heating by $\vec{E}_{0}$ alone cannot
account in full measure for such huge temperature elevations. A
strong correlation between AEH and $E_0$, as well as other
physical arguments, shows that average heating by FB-generated
turbulent electric fields causes AEH
\citep{StMaLaher:85,Robinson:Towards86,Robinson:92,Providakes:88,StMaurice:Unified87,
St.-Maurice:90b,DimantMilikh:JGR03}. AEH occurs largely because
the turbulent electrostatic field,
$\delta\vec{E}=-\nabla\delta\Phi$, has a small component
$\delta\vec{E}_{||}$ parallel to the geomagnetic field
$\vec{B}_0$
\citep{StMaLaher:85,Providakes:88,DimantMilikh:JGR03,MilikhDimant:JGR03,Bahcivan:Parallel2006}.
The importance of $\delta\vec{E}_{||}$ makes the entire process
fully 3-D.

Anomalous electron heating can modify ionospheric conductances
and hence affect the coupling between the magnetosphere and
ionosphere. Any electron heating directly affects the
temperature-dependent electron-neutral collision frequency and,
hence, the electron part of the Pedersen conductivity. This
part, however, is usually small compared to the electron Hall
and ion Pedersen conductivities. However, AEH causes a gradual
elevation of the mean plasma density within the anomalously
heated regions by reducing the local plasma recombination rate
\citep{Gurevich:78,St.-Maurice:90b,DimantMilikh:JGR03,MilikhGoncharenko:Anom2006}.
The AEH-induced plasma density elevations increase all
conductivities in proportion. However, this mechanism requires
tens of seconds or even minutes because of the slow development
of the ionization-recombination equilibrium. If $\vec{E}_0$
changes faster than the characteristic recombination timescale
then its time-averaged effect on density will be smoothed and
reduced.

\begin{figure}
\noindent\includegraphics[width=21pc]{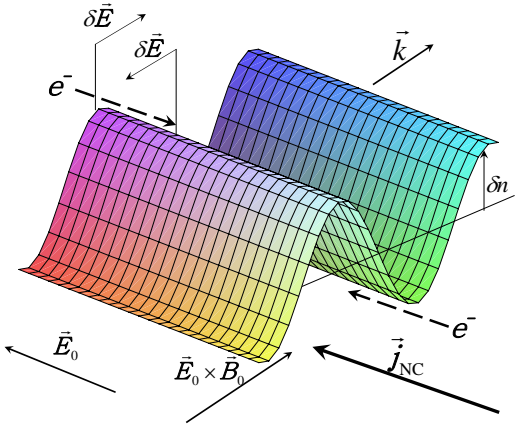}
\caption{Formation of a net non-linear current (NC) at a given wave of
plasma compression/decompression with the wavevector $\vec{k}$ in the
$\vec{E}_0\times \vec{B}_0$-direction. The wave electrostatic field,
$\delta \vec{E} \| \vec{k}$, has opposite directions in the plasma
density maxima ($\delta n>0$) and minima ($\delta n<0$), resulting
in the oppositely directed $\delta\vec{E}\times \vec{B}_0$ drifts
of magnetized electrons. More negatively charged particles move in
the $-\vec{E}_0$ than those in the opposite direction, resulting in
formation of the net positive current, $\vec{j}_{\mathrm{NC}}$,
parallel to $\vec{E}_0$.}
\label{Fig:NC_sketch}
\end{figure}

Plasma turbulence, however, can directly modify local
ionospheric conductivities via a wave-induced non-linear
current (NC) associated with plasma density irregularities
\citep{RogisterJamin:1975,Oppenheim:Evidence97,Buchert:Effect2006}.
The physical nature of the NC is explained in
Fig.~\ref{Fig:NC_sketch} for magnetized electrons and
unmagnetized ions. The wave field, $\delta\vec{E}$, has
different signs in the wave density maxima and minima, so that
the corresponding $\delta\vec{E}\times\vec{B}_{0}$-drifts of
magnetized electrons have opposite directions. As a result,
more electrons drift along the maxima than in the opposite
direction along the minima, producing a net non-linear current
density, $\vec{j}_{\mathrm{NC}}$. At higher \emph{E}-region
altitudes, partially magnetized ions can also contribute to
this current. Relative density perturbations in saturated FB
turbulence may reach at most tens percent, while the rms
turbulent field, $\langle\delta\vec{E}^{2}\rangle^{1/2}$, is
comparable to $E_{0}$ (the angular brackets here and below
denote spatial-temporal averaging). As a result, the total NC,
considered as a plasma response to the external electric field
$\vec{E}_{0}$, amounts to only a fraction of the regular
electrojet Hall current. However, in most of the electrojet the
NC is directed largely parallel to $\vec{E}_{0}$, so that it
may increase significantly the much smaller Pedersen
conductivity. This is critically important because the Pedersen
conductivity allows the MI field-aligned currents to close and
dissipate energy. The combined effect of the NC and AEH makes
the ionosphere less resistive. The anomalous conductance may,
at least partially, account for systematic overestimates of the
total cross-polar cap potential in global MHD models that
employ laminar conductivities
\citep[e.g.,][]{winglee:1997,raeder:1998,raeder.monograph:2001,Siscoe:Hill2002,
Ober:Testing2003,
Merkin:Global2005,Merkin:Anomalous2005,Merkin:Predicting2007,guild:2008a,wang.h:2008}.

Furthermore, the NC and AEH are intrinsically related. As
discussed in this paper, the work by the external DC electric
field on the total current equals the total field energy input
to ionosphere (total Joule heating), including that responsible
for AEH. On the other hand, the total frictional heating of
electrons and ions effectively increases wave dissipation and
reduces the turbulence intensity, thus affecting the NC.

The process of anomalous energy deposition from the
magnetosphere to ionosphere has been recently studied by
\citet{Buchert:Effect2006}. They showed that the apparent
average turbulent energy deposition per unit volume and time,
$\langle\delta\vec{E}\cdot\delta\vec{j}\rangle$, equals zero
(here $\delta\vec{j}$ is the fluctuation of the total current
density $\vec{j}$; a number of notations in this paper differ
from those of \citet{Buchert:Effect2006}). The authors deduced
that the total energy per unit volume and time lost by the
external field $\vec{E}_{0}$ to particles equals the total work
performed by this field on the total average current,
$\vec{j}=\vec{j}_{0}+\vec{j}_{\mathrm{NC}}$ ($\vec{j}_{0}$ is
the laminar current density). They suggested that the actual
energy input for the instability development and
turbulence-induced energy losses equals
$\vec{E}_{0}\cdot\vec{j}_{\mathrm{NC}}$. Their calculations
supported this remarkable conjecture, but provided no complete
proof since their quantitative analysis was based on an
oversimplified and restrictive model. First and most
importantly, they studied only 2-D turbulence in a
perpendicular to $\vec{B}_{0}$ plane. The authors did not state
the 2-D restriction explicitly, but it becomes evident from
their Eqs.~(17) and (18) which effectively exclude
$\vec{k}_{\parallel}$. Due to this, their treatment leaves out
the key 3-D energy conversion process primarily responsible for
AEH. Secondly, they employed the simplest two-fluid model with
fully demagnetized ions. This widely used approximation is good
for most of the electrojet but fails at its top part, right
where the Pedersen conductivity reaches its maximum. Thirdly,
their calculations of NC and other average terms were fully
based on a quasilinear, narrowband approximation for FB waves.
In this approximation, turbulence consists of relatively
small-amplitude waves which are described by linear relations
with narrowband wave frequencies. In many cases, this is a
reasonable approach, although its applicability is less
justified for the driving electric field well above the
instability threshold, when the effect of AEH is especially
strong.

These restrictions and some inconclusiveness associated with
them raise a number of important questions. First and foremost,
is the conjecture about $\vec{E}_{0}\cdot\vec{j}_{\mathrm{NC}}$
as the energy input for plasma turbulence really true? If so,
then this fundamental fact should follow directly from the
first principles and be universally applicable. In particular,
it has to be valid for arbitrarily magnetized particles and
strongly non-linear processes. Then, is it possible to deduce
this fact from a general viewpoint with no or minimal
approximations? Second, presuming that
$\vec{E}_{0}\cdot\vec{j}_{\mathrm{NC}}$ is the correct energy
input, then how is it compatible with the 3-D effect of AEH
largely associated with small $\delta\vec{E}_{\parallel}$? This
question arises because $\vec{j}_{\mathrm{NC}}$, as well as
$\vec{E}_{0}$, lies almost precisely in the plane perpendicular
to $\vec{B}_{0}$ with no average contribution from
$\vec{k}_{\parallel}$. Recently, we performed a number of 2-D
and 3-D\ particle-in-cell (PIC) simulations in big periodic
boxes with unprecedentedly dense meshes and large numbers of
PIC particles \citep{Oppenheim:Fully2011}. Comparing the 2-D
and 3-D results for the same background parameters, we have
obtained compelling evidence that anomalous electron heating
related to plasma structuring along $\vec{B}_{0}$ does exist.
Then, how could essentially 3-D heating originate from the 2-D
energy input? Further, how this energy input is distributed
between various groups of particles? This is important for
making accurate estimates of anomalous heating for different
particle groups. During strong magnetospheric events, what
feedback of developed \emph{E}-region turbulence on global MHD
behavior of the magnetosphere might be expected? How to
quantify this effect for including it in global MHD codes
intended for space weather predictions? Finally, what channels
provide the corresponding energy flow between the magnetosphere
and \emph{E}/\emph{D}-region ionosphere?

In this paper, we address these issues and create a rigorous
basis for calculating anomalous conductivities in the companion
paper \citep{Dimant:Magnetosphere2011}. We start by confirming
from first principles that fully saturated turbulence does
yield $\langle\delta\vec{E}\cdot\delta\vec{j}\rangle=0$, then
confirm the \citet{Buchert:Effect2006} deduction regarding the
turbulent energy input and establish its universal validity. In
order to quantitatively develop a 3-D model of AEH and resolve
the apparent contradiction between this interpretation and the
2-D nature of the energy input, we perform specific
calculations for the case of arbitrary particle magnetization,
using a quasilinear approximation. We calculate the non-linear
current, total energy input, and partial average frictional
heating sources for both electrons and ions in terms of a given
spectrum of density irregularities. We show that the major
quantitative difference between 2-D and 3-D developed
turbulence lies in the magnitude of density perturbations.
These perturbations and the non-linear current proportional to
them are noticeably larger in 3-D than in 2-D. This difference
explains the larger energy input in 3-D and is responsible for
AEH caused by turbulent fields, thus resolving the
above-mentioned contradiction.

The paper is organized as follows. In Sect.~\ref{General
consideration}, using only first principles with no
approximations, we confirm for the general case the
\citet{Buchert:Effect2006} findings regarding the energy input.
In Sect.~\ref{Partial energy deposit: quasilinear approach}, we
develop a quasilinear approach, similar to that of
\citet{Buchert:Effect2006}, but for the general 3-D case of
arbitrarily magnetized particles. This allows us to calculates
partial non-linear currents, relevant energy inputs, and
frictional heating sources. In Sect.~\ref{Global Energy Flow},
we discuss global energy flow between the magnetosphere and
ionosphere. In the appendix, we check the validity of the
conventional electrostatic approximation for lower-ionosphere
wave processes.

\section{Energy Conversion: First Principle Consideration\label{General consideration}}

In this section, we derive general relations regarding the
average energy input in quasi-periodic systems and show how the
\citet{Buchert:Effect2006} deductions follow from fundamental
electrodynamic and plasma kinetics principles with no
approximations like electrostatics, quasi-neutrality,
fluid-model description, etc.

First, we consider the evolution of the field energy in plasmas
by considering the exact electrodynamics, starting with
Ampere's and Faraday's laws,
\begin{subequations}
\label{Maxwell's}%
\begin{align}
\partial_{t}\vec{E}  &  =c^{2}\left(  \nabla\times\vec{B}\right)  -\frac
{\vec{j}}{\varepsilon_{0}},\label{Ampere's}\\
\partial_{t}\vec{B}  &  =-\nabla\times\vec{E},%
\label{Faraday's}
\end{align}
\end{subequations}
where $\varepsilon_{0}$ is the permittivity of free space, $c$
is the speed of light in vacuum, $\vec{E}$ and $\vec{B}$ are
the electric field and magnetic induction, and $\vec{j}$ is the
total current density.

Taking scalar products of Eq.~(\ref{Ampere's}) with $\varepsilon_{0}\vec{E}$,
Eq.~(\ref{Faraday's}) with $\varepsilon_{0}c^{2}\vec{B}
$ and adding the results, we obtain the standard energy balance
equation (aka Poynting's theorem):
\begin{equation}
\partial_{t}U+\nabla\cdot\vec{S}=-\vec{E}\cdot\vec{j}, \label{Poynting's}%
\end{equation}
where
\begin{equation}
U=\frac{\varepsilon_{0}%
}{2}\left(  E^{2}+c^{2}B^{2}\right)  ,\qquad\vec{S}\equiv\frac{\vec{E}%
\times\vec{B}}{\mu_{0}}
\label{US}
\end{equation}
are the field energy density and the corresponding flux (the Poynting vector),
respectively; $\mu_{0}=(\varepsilon_{0}c^{2})^{-1}$ is the permeability of
free space.

Now we need to find $\vec{j}$ from the plasma. The dynamics of
individual particles of type $s$, such as electrons or ions
($s=e,i$), is accurately described by Boltzmann's kinetic
equation, which can be written in the 6-D divergence form as
\begin{equation}
\partial_{t}f_{s}+\nabla\cdot\left(  \vec{v}_{s}f_{s}\right)
+\partial_{\vec{v}_{s}}\cdot\left[  \frac{q_{s}}{m_{s}}(\vec
{E}+\vec{v}_{s}\times\vec{B})f_{s}\right]  =S_{s}.
\label{Kinetic_alpha}%
\end{equation}
Here $\vec{v}_{s}$ is the kinetic velocity of the
particles-$s$, while $f_{s}(\vec{r},t,\vec{v}_{s})$ is their
single-particle velocity distribution function normalized to
the particle-$s$ density, $n_{s}\equiv\int f_{s}d^{3}v_{s}$
(the integration here and below is performed over the entire
3-D velocity space); $q_{s}$ and $m_{s}$ are the particle
charge and mass; $S_{s}$ is the collisional operator which
includes particle-$s$ collisions and can also include the
ionization sources and recombination losses; for simplicity, we
disregard effects of gravity. Multiplying
Eq.~(\ref{Kinetic_alpha}) by $m_{s}v_{s}^{2}/2$, integrating
over the velocity space, and adding the results for all plasma
particles, we obtain the energy balance relation for plasma,
\begin{equation}
\partial_{t}\sum_{s}\mathcal{E}_{s}+\nabla\cdot\sum_{s}\vec
{K}_{s}=\vec{E}\cdot\vec{j}+\sum_{s}L_{s}.
\label{particle_energy _conserva}%
\end{equation}
Here $\mathcal{E}_{s}\equiv\int(m_{s}v_{s}^{2}/2)\ f_{s
}d^{3}v_{s}\ $and $\vec{K}_{s}\equiv\int(m_{s}v_{s}%
^{2}/2)\vec{v}_{s}f_{s}d^{3}v_{s}$ are the average particle
energy and energy-flux densities, respectively; $L_{s}\equiv
\int(m_{s}v_{s}^{2}/2)\allowbreak S_{s}d^{3}v_{s}$
combines all collisional energy gains and losses; $\vec{j}\equiv\sum_{s
}q_{s}n_{s}\vec{V}_{s}$, is the total current density, the same
as in Eq.~(\ref{Poynting's}); $\vec{V}_{s}\equiv\int\vec{v}_{s
}f_{s}d^{3}v_{s}/n_{s}$ is the particle-$s$ mean fluid
velocity. In the general case, the particle energy density $\mathcal{E}%
_{s}$ combines the mean thermal energy, $3n_{s}T_{s}/2$, where
$T_{s}\equiv\int[m_{s}(v_{s}-\vec{V}_{s})^{2}/3]f_{s
}d^{3}v_{s}/n_{s}$ is the effective particle temperature, with the
kinetic energy density of the mean particle flow, $n_{s}m_{s
}V_{s}^{2}/2$.

Comparison of Eqs.~(\ref{Poynting's}) and (\ref{particle_energy
_conserva}) shows that $\vec{E}\cdot\vec{j}$ is the total
energy input from the fields to particles per unit volume and
time. The energy deposited in an individual group of particles
at a given location may be then slightly redistributed via
Coulomb collisions and transported to other locations.
Eventually, this energy  becomes lost to the abundant neutral
atmosphere via predominantly inelastic plasma-neutral
collisions.

General Eqs.~(\ref{Poynting's}) and (\ref{particle_energy
_conserva}) apply to all plasma processes, linear or
non-linear. To specifically discuss turbulent processes, we
need to separate them from the slowly evolving and large-scale
regular background structures and processes. This can be done
by using a conventional two-scale approach in which macroscopic
background structures and processes are presumed to have much
longer characteristic spatial and temporal scales than does the
turbulence. The physical conditions in the \emph{E}-region
plasma make this procedure more applicable to plasma
irregularities generated by local instability mechanisms, such
as the Farley-Buneman (FB) and thermal-driven instabilities
\citep{Dimant:Physical97,Kagan:thermal00,DimOppen2004:ionthermal1}
than for relatively large-scale irregularities generated by
plasma gradients.

We assume that plasma turbulence generated by instabilities
consists mostly of waves whose wavelengths $\lambda_{i}$ in
each direction $i$ are much smaller than the corresponding
typical scales of spatial variation of the macroscopic
background parameters, $\Lambda_{i}$. Then, for any given
location $\vec{r}$, we can build an imaginary rectilinear box
centered around $\vec{r}$ with the sizes $L_{i}$ that satisfy
the conditions $\lambda_{i}\ll L_{i}\ll\Lambda_{i}$. The
characteristic scales in different directions $i$ can differ
dramatically. For example, the wavelengths of nearly
field-aligned irregularities parallel to $\vec{B}_{0}$ are
generally orders of magnitude larger that those perpendicular
to $\vec{B}_{0}$, so that the corresponding box sizes would
also be vastly different. Within the imaginary box, we can
extend the background parameters from the box center $\vec{r}$
uniformly to the entire box and impose periodic boundary
conditions. This approach is widely employed in computer
simulations.

We can similarly describe the temporal evolution of the process
by presuming long-lasting non-linearly saturated turbulence in
a quasi-stationary background. This implies a continuous
driving of instability and allows one to introduce a
quasi-period $T$, analogous to $L_{i}$, which is much longer
than typical timescales of turbulent variations but is much
shorter than a characteristic timescale of slow evolution of
the macroscopic background.

To implement the two-scale procedure, we introduce the
following convention. We denote the short-scale and fast
periodic variables for turbulent processes by
$\vec{x}=(x_{1},x_{2},x_{3})$ and $t$, while denoting the
corresponding coordinates and time of large-scale and slow
background variations by $\vec{r}$ and $\tau$, respectively.
With respect to the periodic coordinates $\vec{x}$ and time
$t$, we define a spatial-temporal average as
\begin{equation}
\left\langle \cdots\right\rangle \equiv\frac{1}{L_{1}L_{2}L_{3}T}\!\int
_{-L_{1}/2}^{L_{1}/2}\!\!\!dx_{1}\!\!\int_{-L_{2}/2}^{L_{2}/2}\!\!\!dx_{2}\!
\int_{-L_{3}
/2}^{L_{3}/2}\!\!\!dx_{3}\!\int_{\tau-T/2}^{\tau+T/2}
\!\!\!\!\!\!\!\!\left(\cdots\right)
 dt.
\label{spatial_average}%
\end{equation}
In general, such averages can remain functions of the slow variables $\vec{r}$ and
$\tau$. This description of turbulence does not require the ensemble
averaging employed by \citet{Buchert:Effect2006}.

Having implemented this procedure, we split the quasi-periodic
fields and currents into their $\vec{r}$,$\tau$-dependent
average parts and the corresponding local periodic
perturbations,
\begin{equation}
\vec{E}=\langle\vec{E}\rangle+\delta\vec{E},\qquad\vec{B}=\langle\vec
{B}\rangle+\delta\vec{B},\qquad\vec{j}=\langle\vec{j}\rangle+\delta\vec{j},
\label{perturbations}%
\end{equation}
Using $\langle\delta\vec{E}\rangle=\langle\delta\vec{j}\rangle=0$, we obtain
for the average energy deposition term%
\begin{equation}
\langle\vec{E}\cdot\vec{j}\rangle=\langle\vec{E}\rangle\cdot\langle\vec
{j}\rangle+\langle\delta\vec{E}\cdot\delta\vec{j}\rangle. \label{split_EJ}
\end{equation}
One might naively interpret the first term on the right-hand
side (RHS) of Eq.~(\ref{split_EJ}) as the energy input from the
background fields to the undisturbed plasma, while the second
term as the corresponding average contribution from the
turbulent fields. As mentioned in the Introduction,
\citet{Buchert:Effect2006} found, using a simplified model,
that $\langle\delta\vec{E}\cdot\delta\vec{j}\rangle$ equals
zero by the following formal mathematical reason. If one
expands this term further in terms of the mean particle fluid
velocities to the lowest-order quadratic non-linearity then the
expected total turbulent frictional heating term,
$\sum_{s}q_{s}n_{s0}\langle\delta\vec{E}\cdot\delta\vec
{V}_{s}\rangle$, turns out to be automatically canceled by a
density perturbation term, $\sum_{s}q_{s}\langle\delta
n_{s}\delta\vec{E} \rangle\cdot\vec{V}_{s0}$ (here $\delta
n_{s}$ and $\delta\vec {V}_{s}$ are the perturbations of the
densities and particle fluid velocities). Right below we show
that for purely periodic and spatially homogeneous turbulence
the equality $\langle\delta\vec{E} \cdot\delta\vec{j}\rangle=0$
is a natural and universal constraint, required merely by the
imposed periodicity. This constraint follows directly from
exact Maxwell's equations without invoking specific plasma
models.

Indeed, Maxwell's equations are linear, allowing the separation
of the average quantities from wave perturbations. The
perturbations depend on the small-scale, quasi-periodic
variables, $\vec{x}$, $t$, and can also have an adiabatically
slow $\vec{r}$, $\tau$-dependence due to background
inhomogeneities and evolution. In principle, the inhomogeneous
background parameters may evolve in such a way that an
instability threshold is crossed, resulting in sudden onset or
disappearance of instability in some locations at certain
moments of time. Such instances, analogous to second-order
phase transitions \citep{LandauStatistical_1_2000}, break the
validity of our two-scale approximation and deserve a special
treatment that lies beyond the framework of this paper. Apart
from these special occasions, we can apply to the perturbations
of the fields and current the same steps that lead to
Eq.~(\ref{Poynting's}). Averaging the result in accord with the
definition of Eq.~(\ref{spatial_average}), we obtain
\begin{equation}
\partial_{\tau}\left[  \frac{\varepsilon_{0}\left(  \left\langle \delta
E^{2}\right\rangle +c^{2}\left\langle \delta B^{2}\right\rangle \right)}
{2}\right]  +\nabla_{\vec{r}}\cdot\frac{\langle\delta\vec{E}\times\delta
\vec{B}\rangle}{\mu_{0}}=-\langle\delta\vec{E}\cdot\delta\vec{j}\rangle.
\label{Poynting's_perturb}%
\end{equation}
Then for saturated turbulence with constant average
characteristics in a strictly periodic box, the left-hand side
(LHS) of Eq.~(\ref{Poynting's_perturb}) disappears, yielding
\begin{equation}
\langle\delta\vec{E}\cdot\delta\vec{j}\rangle=0, \label{so_that_really}%
\end{equation}
regardless of the specific fluid or kinetic models employed for
the plasma description.  For strictly periodic processes, this
result is exact. One can also obtain it directly from the
discrete Fourier harmonics of the electric field and current.
We do this right below and also demonstrate that electrostatic
and quasi-neutral approximations do not lift this exact
electrodynamic constraint.

We start by introducing spatial and temporal Fourier harmonics
of wave perturbations for general, linear or non-linear,
periodic processes. Since we imply saturated turbulence in a
4-D box (3-D space + time) with periodic boundary conditions,
it is logical to employ the 4-D discrete Fourier
transformation. For easy reading, we assign to the Fourier
transforms the notations of the original variables but with
additional subscripts $\vec {k},\omega$. For a scalar or vector
periodic perturbation $\delta F(\vec {x},t)$ as a function of
coordinates $x_{i}$ and time $t$, we define such
transformations within the 4-D box of the corresponding sizes
$L_{i}$ and the quasi-period $T$ as
\begin{subequations}
\label{Fourier_transform}%
\begin{align}
 &\delta F(\vec{x},t)   =\sum_{\vec{k},\omega\neq0}\delta F_{\vec{k},\omega
}\exp[i(\vec{k}\cdot\vec{x}-\omega t)],\label{Function}\\
&\delta F_{\vec{k},\omega}    =\frac{1}{L_{1}L_{2}L_{3}T}\!\!\int\!\!\delta F(\vec
{x},t)\exp\!\left[-i\!\left(\sum_{i=1}^3k_{i}x_{i}-\omega t\right)\!\right]\!\!
d^{3}x_{i}dt.%
\label{Transform}
\end{align}
\end{subequations}
Here the $\vec{k},\omega\neq0$ summation is taken over all
discrete values of $\vec{k}$, $k_{i}=2\pi n_{i}/L_{i}$,
$\omega=2\pi m/T$, with integer $n_{i}$, $m$, while the
integration is performed over the entire 4-D box, as in
Eq.~(\ref{spatial_average}). Since we apply the discrete
Fourier transforms only to the perturbations, we exclude from
the summation over harmonics the average values corresponding
to $\vec{k},\omega =0$. In terms of their Fourier harmonics,
$\delta A_{\vec{k},\omega}$ and $\delta B_{\vec{k},\omega}$,
the spatial-temporal average of any product, scalar or vector,
of two functions $A(\vec{x},t)$ and $B(\vec{x},t)$, is given by
\begin{equation}
\left\langle \delta A(\vec{x},t)\delta B(\vec{x},t)\right\rangle =\sum
_{\vec{k},\omega\neq0}\delta A_{\vec{k},\omega}\delta B_{\vec{k},\omega}%
^{\ast}=\sum_{\vec{k},\omega\neq0}\delta A_{\vec{k},\omega}^{\ast}\delta
B_{\vec{k},\omega}.\label{Average_harmonics}%
\end{equation}
Our current discrete-transform normalization differs from the
continuous one in \citet{Buchert:Effect2006} and allows to
avoid the emergence of extraneous factors like $VT$
($V=L_1L_2L_3$) and $(2\pi)^4$ in the explicit physical
expressions for average quadratically non-linear quantities
\citep[e.g.,][Eqs.~(6)--(8), (19) --(26),
etc.]{Buchert:Effect2006}.

Using Fourier transforms, we now prove
Eq.~(\ref{so_that_really}) in a more direct way. Applying
Eq.~(\ref{Fourier_transform}) to Eq.~(\ref{Maxwell's}), for a
given Fourier harmonic,  we obtain
\begin{equation}
ic^{2}\vec{k}\times\delta\vec{B}_{\vec{k},\omega}=\frac{1}{\varepsilon_{0}%
}\ \delta\vec{j}_{\vec{k},\omega}-i\omega\delta\vec{E}_{\vec{k},\omega},\qquad
i\vec{k}\times\delta\vec{E}_{\vec{k},\omega}=i\omega\delta\vec{B}_{\vec
{k},\omega}. \label{urki}%
\end{equation}
Making a cross-product of Eq.~(\ref{urki}) with $\vec{k}$\ and
using $\vec {k}\cdot\delta\vec{B}_{\vec{k},\omega}=0$, we
express the field perturbations in terms of
$\delta\vec{j}_{\vec{k},\omega}$,
\begin{align}
\delta\vec{B}_{\vec{k},\omega}  &  =\frac{i\vec{k}\times\delta\vec{j}_{\vec
{k},\omega}}{\varepsilon_{0}(k^{2}c^{2}-\omega^{2})},\label{del_B}\\
\delta\vec{E}_{\vec{k},\omega}  &  =\frac{i\left[  \omega^{2}\delta\vec
{j}_{\vec{k},\omega}-c^{2}(\vec{k}\cdot\delta\vec{j}_{\vec{k},\omega})\vec
{k}\right]  }{\varepsilon_{0}\omega(k^{2}c^{2}-\omega^{2})}. \label{del_E}%
\end{align}
In a long-lived quasi-periodic non-linearly saturated state,
all wave frequencies $\omega$ must be real, so that
Eq.~(\ref{del_E}) yields
$\operatorname{Re}(\delta\vec{E}_{\vec{k},\omega}\cdot\delta\vec{j}_{\vec{k}
,\omega}^{\ast})=0$, i.e., Eq.~(\ref{so_that_really}).

General Eq.~(\ref{so_that_really}) represents a strict
constraint that follows from the full electrodynamics of
quasi-periodic processes, but it is not obvious that it should
hold for the electrostatic and quasi-neutrality approximations.
The following demonstrates that these approximations do not
lift this constraint.

Indeed, the electron and ion continuity equations combined
yield $e\partial_t (\delta n_e - \delta n_i) = \nabla
\delta\vec{j}$, so that $ e\omega(\delta n_{i\vec{k},\omega} -
\delta n_{e\vec{k},\omega}) = \vec{k}\cdot
\delta\vec{j}_{\vec{k},\omega} $. Combining this with Poisson's
equation, $\varepsilon_0\nabla\cdot\delta\vec{E}=e(\delta n_e -
\delta n_i)$, we obtain
\begin{equation}
i\vec{k}\cdot\delta\vec{E}_{\vec{k},\omega}=
\frac{\vec{k}\cdot\delta\vec{j}_{\vec{k},\omega}}{\varepsilon_0\omega}.%
\label{proshka1}
\end{equation}
For an electrostatic field, we have
$\delta\vec{E}_{\vec{k},\omega}=|\delta\vec{E}_{i\vec{k},\omega}|(\vec{k}/k)$,
so that Eqs.~(\ref{Average_harmonics}) and (\ref{proshka1})
yield
\begin{equation}
\langle\delta \vec{j}\cdot\delta\vec{E}\rangle = \mathrm{Re}
(\delta \vec{j}_{\vec{k},\omega}\cdot\delta\vec{E}_{\vec{k},\omega}^\ast)
= -\varepsilon_0 |\delta\vec{E}_{\vec{k},\omega}|^2 \,\mathrm{Im}\,\omega = 0.%
\label{ej=0_electrostat}
\end{equation}
If we add quasi-neutrality, $\nabla\cdot \delta\vec{j}=0$,
i.e., $\vec{k}\cdot\delta\vec{j}_{\vec{k},\omega}=0$ for
individual harmonics, then the satisfaction of
Eq.~(\ref{so_that_really}) becomes obvious even without
assumption of real $\omega$,
\begin{equation}
\langle \delta\vec{j}\cdot\delta\vec{E}\rangle =i\sum_{\vec{k},\omega
\neq0}(\vec{k}\cdot  \delta\vec{j}_{\vec{k},\omega})
\delta\Phi_{\vec{k},\omega}^{\ast}=0,%
\label{ej=0_quasineutral}
\end{equation}

Equation (\ref{so_that_really}) does not mean, however, that
developed plasma turbulence makes no contribution to the
average particle heating. This would certainly contradict both
observations and PIC simulations. The paradox can be resolved
as follows. The average electric field is merely the external
field, $\langle\vec{E}\rangle=\vec{E}_{0}$, while $\langle\vec
{j}\rangle\neq\vec{j}_{0}$, where
$\vec{j}_{0}\equiv\sum_{s}q_{s }n_{s0}\vec{V}_{s0}$ is the
undisturbed current density determined by the laminar plasma
response to $\vec{E}_{0}$. In addition to $\vec{j}_{0}$, the
total average current density, $\langle\vec{j}\rangle$,
includes a wave-induced direct NC
\citep{RogisterJamin:1975,Oppenheim:Evidence97,Buchert:Effect2006},
\begin{equation}
\vec{j}^{\mathrm{NC}}=\langle\vec{j}\rangle-\vec{j}_{0}\equiv\sum_{s
}q_{s}\langle\delta n_{s}\delta\vec{V}_{s}\rangle. \label{J_NL}%
\end{equation}
The total average loss of field energy to particles per unit
volume and time is given by
$\langle\vec{E}\cdot\vec{j}\rangle$. The corresponding energy
dissipation in the laminar ionosphere with no instabilities is
given by $\vec{E}_{0}\cdot\vec{j}_{0}$. Then the total energy
dissipation exclusively due to plasma turbulence is
$L_{\mathrm{turb}}\equiv\langle\vec{E}\cdot\vec
{j}\rangle-\vec{E}_{0}\cdot\vec{j}_{0}$. Using
Eqs.~(\ref{so_that_really}) and (\ref{J_NL}), one can easily
establish that
\begin{equation}
L_{\mathrm{turb}}=P_{\mathrm{NC}}\equiv\vec{E}_{0}\cdot\vec{j}_{\mathrm{NC}}.
\label{L_turb}%
\end{equation}
Thus, it is formally the work by the external electric field on
the non-linear current, $P_{\mathrm{NC}}$, rather than
$\langle\delta\vec{E}\cdot\delta \vec{j}\rangle$, that provides
the required turbulent energy deposition for all kinds of
anomalous heating of plasma particles.
\citet{Buchert:Effect2006} showed this for the restricted case
of fluid plasmas with fully unmagnetized ions and a quasilinear
wave description, but did not establish it in the general case.
We have just demonstrated that this fundamental result follows
directly from the general field electrodynamics and no specific
plasma models. By their physical meaning and according to
Eq.~(\ref{L_turb}) each of the two equal quantities
$L_{\mathrm{turb}}$ and $P_{\mathrm{NC}}$ can be named
``turbulent Joule heating.''

Rigorously speaking, exact Eqs.~(\ref{so_that_really}) and
(\ref{L_turb}) apply only to a homogeneous and stationary
background. For mildly inhomogeneous and slowly evolving
background parameters, Eq.~(\ref{Poynting's_perturb}) can be
treated using a regular perturbation technique. The zero-order
approximation, corresponding to a given turbulence level within
an isolated box with a uniform background and periodic boundary
conditions and, hence, not affected by the outside
inhomogeneity, yields Eq.~(\ref{so_that_really}). To reach a
next-order accuracy, one has to establish the zero-order
parameter dependence of non-linearly saturated turbulence
characteristics, using, e.g., a series of computer simulations
with periodic boundary conditions but various background
parameters. Given the large-scale spatial dependence and slow
evolution of the background parameters, one could calculate
then the LHS of Eq.~(\ref{Poynting's_perturb}). This would
yield the first-order, potentially non-zero, values for
$\langle\delta\vec{E}\cdot\delta\vec{j}\rangle$ on the RHS.
Under applicability of our two-scale approach, however, such
possible finite values of $\langle\delta
\vec{E}\cdot\delta\vec{j}\rangle$ will automatically be small
compared to the nearly balancing each other leading terms in
the expanded form of $\langle\delta\vec
{E}\cdot\delta\vec{j}\rangle$ with $\delta\vec{j}=\sum_{s}q_{s}
(\vec{V}_{s0}\delta n+n_{0}\delta\vec{V}_{s}+\delta n\delta\vec
{V}_{s})$, creating only a relatively small mismatch between
$L_{\mathrm{turb}}$ and $P_{\mathrm{NC}}$.

Below we calculate energies deposited among individual particle
groups using an approach similar to that used by
\citet{Buchert:Effect2006}, except that we will apply it to a
system with arbitrarily magnetized electrons and ions. This
applies to all altitudes across the entire \textit{E} region
down to the upper \textit{D} region. As mentioned above,
\citet{Buchert:Effect2006} actually performed a 2-D treatment.
We consider here fully 3-D turbulence, which is crucial for
anomalous electron heating. Note that the total energy input
spent on anomalous heating of plasma particles is less than
$P_{\mathrm{NC}}$ because a fraction of the deposited energy
via collisions goes directly to colliding neutrals and has no
chance to heat the plasma.

\section{Partial Energy Deposit: Quasilinear
Approximation\label{Partial energy deposit: quasilinear
approach}}

In this section, we employ Fourier harmonics in a 3-D periodic
box in order to identify the effect of the parallel turbulent
electric field. Our results apply to all regions that
contribute to the total ionospheric conductances, from the top
electrojet down to the potentially unstable \emph{D}-region
\citep{Kelley:Ionosphere2009}. Similar processes can occur in
other plasma media, like the Solar chromosphere
\citep{Liperovsky:Generation2000,Fontenla:Chromospheric2008,Gogoberidze:Farley2009},
other planetary ionospheres, and laboratory plasma
\citep{Dangelo:1974,John:Observation1975,Koepke:Space2008},
despite the dramatic differences in parameters. Lastly, these
calculations for arbitrarily magnetized plasma serve as an
additional verification of the general relations obtained in
the previous section from first principles.

In this paper, we restrict ionospheric particles to electrons
and a single species of ions, adopting the quasi-neutrality,
$n_{e}\approx n_{i}=n$. Similarly to
\citet{Buchert:Effect2006}, we apply here a quasilinear
approximation, by which we imply that separate Fourier
harmonics of predominantly electrostatic field and plasma
density fluctuations are coupled through simple linear
relations. Using these relations, we calculate then
quadratically non-linear averages in terms of given
spatial-temporal turbulence spectra.

This section is organized as follows. In Sect.~\ref{Dispersion
relation and particle heating: first-order approximation}, we
obtain linear relationships to first-order accuracy and explain
what we mean by the different orders. These first-order
relations provide the linear wave frequencies and relations
between the electrostatic potential and density perturbations.
In Sect.~\ref{non-linear current} we calculate the partial and
total non-linear currents in terms of a given spectrum of
irregularities. In Sect.~\ref{Energy inputs and turbulent
heating}, we calculate partial energy inputs and turbulent
heating of electrons and ions and verify that general
Eqs.~(\ref{so_that_really}) and (\ref{L_turb}) remain in this
approximation exactly valid.

\subsection{First-Order Linear Wave Relations
\label{Dispersion relation and particle heating: first-order approximation}}

\citet{Fejer:Theory84} studied the linear theory of collisional
waves for strongly magnetized electrons and arbitrarily
magnetized ions. On the other hand, \citet{Buchert:Effect2006}
assumed arbitrarily magnetized electrons but unmagnetized ions.
Since no one has published the general 3-D linear relations
that cover all cases, we do this here and a more general
version in the appendix of the companion paper
\citep{Dimant:Magnetosphere2011}. In reasonable agreement with
our PIC simulations, we assume that most of developed
turbulence lies in the long-wavelength, low-frequency range of
$kl_{i}\ll 1$ and $\omega \ll \nu_{i}$, where $\omega$ is the
wave frequency and $l_{i}$ is the mean free path of ions with
respect to dominant ion-neutral collisions. This allows us to
employ a two-fluid, as opposed to kinetic, model and order
various terms in the momentum equations with respect to the
small parameters $kl_{i}$ and $\omega / \nu_{i}$. In this
ordering, particle inertia and pressure gradients are
second-order effects and can be neglected. This first-order
approximation yields a dispersion relation and the
corresponding relation between fluctuations of the plasma
density and electrostatic potential. These relations are common
for all \emph{E}/\emph{D}-region plasma instabilities
\citep{DimOppen2004:ionthermal1}. The neglected second-order
corrections are crucial for the linear wave dissipation and
instability driving, but they are of less importance to the
spatially/temporally averaged energy transfer and plasma
heating. To the second-order accuracy, the general linear wave
theory for arbitrarily magnetized plasmas is developed in the
appendix of \citet{Dimant:Magnetosphere2011}.

Though we discuss here the linear wave relationships, that does
not mean that we consider the linear stage of instability. On
the contrary, we assume a fully developed and non-linearly
saturated turbulence in which the linear wave growth is
balanced by non-linearities. However, we presume that these
non-linearities only weakly modify the linear wave
relationships, so that we include all non-linear terms, along
with the instability driving or damping terms, into the
second-order corrections and neglect their feedback on the
first-order relations.

Under the first-order approximation, each group of particles has only two
balancing forces: the Lorentz force and resistive collisional friction,
\begin{equation}
q_s(\vec{E}+\vec{V}_{s}\times\vec{B})    = m_{s}\nu_{s}\vec{V}_{s},%
\label{Lorentz}%
\end{equation}
where $s=e,i$; we presume ${V}_{s}$ lies a neutral frame of
reference and an undisturbed magnetic field,
$\vec{B}=\vec{B}_{0}$. Introducing the conventional
magnetization parameters, $\kappa_{s}\equiv\Omega_{s}/\nu_{s}$,
where $\Omega_{s}=|q_s|B/m_s$ are the electron and ion
gyrofrequencies, we obtain from
Eq.~(\ref{Lorentz})%
\begin{equation}
\vec{V}_{e\parallel}=-\ \frac{\kappa_{e}\vec{E}_{\parallel}}{B},\qquad\vec
{V}_{i\parallel}=\frac{\kappa_{i}\vec{E}_{\parallel}}{B}, \label{V_e,i_II}%
\end{equation}%
\begin{equation}
\vec{V}_{e\perp}=\frac{\kappa_{e}[-\vec{E}_{\perp}+\kappa_{e}(\vec{E}%
\times\hat{b})]}{\left(  1+\kappa_{e}^{2}\right)  B},\ \ \ \vec{V}_{i\perp
}=\frac{\kappa_{i}[\vec{E}_{\perp}+\kappa_{i}(\vec{E}\times\hat{b})]}{\left(
1+\kappa_{i}^{2}\right)  B}, \label{V_e,i}%
\end{equation}
with the subscripts $\parallel,\perp$ denoting the components parallel and
perpendicular to $\vec{B}$, respectively; here $B\equiv|\vec{B}|$ and $\hat
{b}\equiv\vec{B}/B$.

Further, we calculate the relative mean velocity between
electrons and ions, $\vec{U}\equiv\vec{V}_{e}-\vec{V}_{i}$,
which plays an important role in many relations.
Equations~(\ref{V_e,i_II}) and (\ref{V_e,i}) give
\begin{subequations}
\label{UUU}%
\begin{align}
\vec{U}_{\parallel}  &  =-\ \frac{\left(  \kappa_{e}+\kappa_{i}\right)
\vec{E}_{\parallel}}{B},\label{U_II}\\
\vec{U}_{\perp}  &  =\frac{\left(  \kappa_{e}+\kappa_{i}\right)  [\left(
\kappa_{e}-\kappa_{i}\right)  (\vec{E}\times\hat{b})-\left(  1+\kappa
_{i}\kappa_{e}\right)  \vec{E}_{\perp}]}{\left(  1+\kappa_{e}^{2}\right)
\left(  1+\kappa_{i}^{2}\right)  B}. \label{U_perp}%
\end{align}
\end{subequations}
Reversing Eq.~(\ref{UUU}), we obtain
\begin{eqnarray}
\frac{\vec{E}_{\parallel}}{B} & = &-\ \frac{\vec{U}_{\parallel}}{\kappa_{e}
+\kappa_{i}},\nonumber
\\
\frac{\vec{E}_{\perp}}{B} & = &-\,\frac{\left(  \kappa_{e}
-\kappa_{i}\right)  (\vec{U}\times\hat{b})+\left(
1+\kappa_{i}\kappa _{e}\right)
\vec{U}_{\perp}}{\kappa_{i}+\kappa_{e}}, \label{ExB_via_U}
\end{eqnarray}
so that Eqs.~(\ref{V_e,i_II}) and (\ref{V_e,i}) yield the following mean drift
velocities of electrons and ions in terms of their relative velocity:%
\begin{equation}
\vec{V}_{e\parallel}=\frac{\kappa_{e}\vec{U}_{\parallel}}{\kappa_{e}%
+\kappa_{i}},\qquad\vec{V}_{i\parallel}=-\ \frac{\kappa_{i}\vec{U}_{\parallel
}}{\kappa_{e}+\kappa_{i}}, \label{V_II_via_U}%
\end{equation}
\begin{subequations}
\label{V_perp_via_U}%
\begin{align}
\vec{V}_{e\perp}  &  =\frac{\kappa_{e}[\vec{U}_{\perp}-\kappa_{i}(\vec
{U}\times\hat{b})]}{\kappa_{e}+\kappa_{i}},\label{V_e_perp_via_U}\\
\vec{V}_{i\perp}  &  =-\ \frac{\kappa_{i}[\vec{U}_{\perp}+\kappa_{e}(\vec
{U}\times\hat{b})]}{\kappa_{e}+\kappa_{i}}. \label{V_i_perp_via_U}%
\end{align}
\end{subequations}
All these linear expressions can be applied separately to the undisturbed
quantities and wave perturbations.

Now we consider wave perturbations of the field, plasma
density, and fluid velocities and apply to them the discrete
Fourier transforms introduced by Eq.~(\ref{Fourier_transform}).
Note that the neglect of non-linear terms with respect to wave
perturbations leads to a unique $\vec{k}$-dependence of the
wave frequency, $\omega=\omega _{\vec{k}}(\vec{k})$, via the
corresponding first-order linear dispersion relation. Such
unique dependence may break the presumed discreteness of the
wave frequency, $\omega=2\pi m/T$. However, the quasi-period
$T$ can always be chosen sufficiently long so that the interval
between adjacent discrete frequencies, $\Delta\omega=2\pi/T$,
becomes much less than the finite spectral width around
$\omega=\omega_{\vec{k}}$; see also the discussion below in the
paragraph following Eq.~(\ref{for_kappa_e>>1}).

Expressing velocity perturbations,
$\delta\vec{V}_{e,i\vec{k},\omega}$, in terms of the
electrostatic potential perturbations,
$\delta\Phi_{\vec{k},\omega}$, via
$\delta\vec{E}_{\vec{k},\omega
}=-i\vec{k}\delta\Phi_{\vec{k},\omega}$, we obtain from
Eqs.~(\ref{V_e,i_II}) and (\ref{V_e,i}):
\begin{subequations}
\label{delta_Vs}%
\begin{align}
\delta\vec{V}_{e\parallel\vec{k},\omega}  &  =i\ \frac{\kappa_{e}\vec
{k}_{\parallel}\delta\Phi_{\vec{k},\omega}}{B},\qquad\delta\vec{V}%
_{i\parallel\vec{k},\omega}=-i\ \frac{\kappa_{i}\vec{k}_{\parallel}\delta
\Phi_{\vec{k},\omega}}{B},\label{delta_V_II}\\
\delta\vec{V}_{e\perp\vec{k},\omega}  &  =-i\ \frac{\kappa_{e}[-\vec{k}%
_{\perp}+\kappa_{e}(\vec{k}_{\perp}\times\hat{b})]\delta\Phi_{\vec{k},\omega}%
}{\left(  1+\kappa_{e}^{2}\right)  B},\label{delta_V_perp_e}\\
\delta\vec{V}_{i\perp\vec{k},\omega}  &  =-i\ \frac{\kappa_{i}[\vec{k}_{\perp
}+\kappa_{i}(\vec{k}_{\perp}\times\hat{b})]\delta\Phi_{\vec{k},\omega}%
}{(1+\kappa_{i}^{2})B}. \label{delta_V_perp_i}%
\end{align}
\end{subequations}
Writing the quasi-neutral continuity equations as
\begin{equation}
\partial_{t}n+\nabla\cdot(n\vec{V}_{i})=0,\qquad\nabla\cdot(n\vec{U})=0,
\label{conti}%
\end{equation}
where we neglected wave variations of ionization-recombination
balance, and linearizing them with respect to  density
perturbations, $\delta n_{\vec{k},\omega}$, we obtain
\begin{equation}
\frac{\delta n_{\vec{k},\omega}}{n_{0}}=\frac{\vec{k}\cdot\delta\vec{V}%
_{i\vec{k},\omega}}{\Omega_{\vec{k}}^{(i)}}=-\ \frac{\vec{k}\cdot\delta\vec
{U}_{\vec{k},\omega}}{\vec{k}\cdot\vec{U}_{0}}. \label{del_n_contin}%
\end{equation}
Here
$\Omega_{\vec{k}}^{(i)}\equiv\omega-\vec{k}\cdot\vec{V}_{i0}$
is the linear wave frequency in the ion frame;
$\delta\vec{U}_{\vec{k},\omega}=\delta\vec
{V}_{e\vec{k},\omega}-\delta\vec{V}_{i\vec{k},\omega}$, and,
according to Eq.~(\ref{UUU}),
\begin{equation}
\vec{U}_{0}=\frac{\left(  \kappa_{i}+\kappa_{e}\right)  [\left(  \kappa
_{e}-\kappa_{i}\right)  (\vec{E}_{0}\times\hat{b})-\left(  1+\kappa_{i}%
\kappa_{e}\right)  \vec{E}_{0}]}{\left(  1+\kappa_{e}^{2}\right)
(1+\kappa_{i}^{2})B}. \label{U_0}%
\end{equation}
Using Eq.~(\ref{delta_Vs}), we obtain
\begin{subequations}
\label{k_Vushki}%
\begin{align}
\vec{k}\cdot\delta\vec{V}_{i\vec{k},\omega}  &  =-i\kappa_{i}\left(
\frac{k_{\perp}^{2}}{1+\kappa_{i}^{2}}+k_{\parallel}^{2}\right)  \frac
{\delta\Phi_{\vec{k},\omega}}{B},\label{k_Vushki_i}\\
\vec{k}\cdot\delta\vec{V}_{e\vec{k},\omega}  &  =i\kappa_{e}\left(
\frac{k_{\perp}^{2}}{1+\kappa_{e}^{2}}+k_{\parallel}^{2}\right)  \frac
{\delta\Phi_{\vec{k},\omega}}{B}, \label{k_Vushki_e}%
\end{align}
\end{subequations}
so that
\begin{equation}
\vec{k}\cdot\delta\vec{U}_{\vec{k},\omega}=\frac{i\kappa_{i}\kappa_{e}\left(
\kappa_{e}+\kappa_{i}\right)  (1+\psi_{\vec{k}})k_{\perp}^{2}\delta\Phi
_{\vec{k},\omega}}{\left(  1+\kappa_{e}^{2}\right)  (1+\kappa_{i}^{2})B},
\label{dul_ka}%
\end{equation}
where
\begin{subequations}
\label{psi}%
\begin{align}
\psi_{\vec{k}}  &  \equiv\psi_{\perp}\left[  1+(1+\kappa_{e}^{2})(1+\kappa
_{i}^{2})\frac{k_{\parallel}^{2}}{k_{\perp}^{2}}\right]  ,\label{psi_k}\\
\psi_{\perp}  &  \equiv\frac{1}{\kappa_{i}\kappa_{e}}=\frac{\nu_{e}\nu_{i}%
}{\Omega_{e}\Omega_{i}}. \label{psi_perp}%
\end{align}
\end{subequations}
The 2-D parameter $\psi_{\perp}$ is conventional. The newly
defined 3-D parameter $\psi_{\vec{k}}$ generalizes the
traditional 3-D parameter $\psi
=\psi_{\perp}(1+\kappa_{e}^{2}k_{\parallel}^{2}/k_{\perp}^{2})$
originally introduced for magnetized electrons,
$\kappa_{e}^{2}\gg1$, and unmagnetized ions,
$\kappa_{i}^{2}\ll1$ \citep[e.g.,][]{Farley:96}. In a more
general case of $\kappa_{e}^{2}\gg1$ but arbitrary
$\kappa_{i}$, the parameter $\psi_{\vec{k}}$ replaces the
product $(1+\kappa_{i}^{2})\psi$ \citep{Fejer:Theory84}. In the
lower ionosphere, the difference
between the two is negligible, $(1+\kappa_{i}^{2})\psi-\psi_{\vec{k}}%
=\kappa_{i}^{2}\psi_{\perp}=\Theta_{0}^{2}\equiv
m_{e}\nu_{e}/(m_{i}\nu _{i})\simeq1.8\times10^{-4}$
\citep{DimantMilikh:JGR03,DimOppen2004:ionthermal1}.

Using Eqs.~(\ref{k_Vushki}) and (\ref{dul_ka}), we obtain from
Eq.~(\ref{del_n_contin}) the first-order relation between the linear
fluctuations of the density and electrostatic potential,%
\begin{equation}
\delta\Phi_{\vec{k},\omega}=i\ \frac{(1+\kappa_{e}^{2})(1+\kappa_{i}%
^{2})B(\vec{k}\cdot\vec{U}_{0})}{\kappa_{i}\kappa_{e}\left(  \kappa_{e}%
+\kappa_{i}\right)  (1+\psi_{\vec{k}})k_{\perp}^{2}}\left(  \frac{\delta
n_{\vec{k},\omega}}{n_{0}}\right)  , \label{Phi->del_n_symmetric}%
\end{equation}
as well as the expression for the first-order wave frequency in the ion frame,
\begin{equation}
\Omega_{\vec{k}}^{(i)}=\frac{(1+\kappa
_{e}^{2})[1+(1+\kappa_{i}^{2})k_{\parallel}^{2}/k_{\perp}^{2}](\vec{k}%
\cdot\vec{U}_{0})}{\kappa_{e}\left(  \kappa_{e}+\kappa_{i}\right)
(1+\psi_{\vec{k}})}. \label{Omega_K}%
\end{equation}
In the neutral frame, the corresponding frequency, $\omega_{\vec{k}%
}=\Omega_{\vec{k}}^{(i)}+\vec{k}\cdot\vec{V}_{i0}$, is
\begin{equation}
\omega_{\vec{k}}=\frac{(\kappa_{e}-\kappa_{i})(\vec{k}\cdot\vec{U}_{0}%
)}{(\kappa_{e}+\kappa_{i})(1+\psi_{\vec{k}})}-\frac{\kappa_{i}\kappa_{e}%
\vec{k}\cdot(\vec{U}_{0}\times\hat{b})}{\kappa_{e}+\kappa_{i}}.
\label{omega_k}%
\end{equation}
When obtaining these expressions, we used the easily derived relation%
\begin{equation}
\left(  1+\kappa_{i}\kappa_{e}\right)  k_{\perp}^{2}+(1+\kappa_{e}%
^{2})(1+\kappa_{i}^{2})k_{\parallel}^{2}=\kappa_{i}\kappa_{e}k_{\perp}%
^{2}(1+\psi_{\vec{k}}). \label{used_here}%
\end{equation}
As might be expected for arbitrarily magnetized electrons and
ions, Eqs.~(\ref{Phi->del_n_symmetric}) and (\ref{omega_k}) are
symmetric with respect to the interchange between electrons and
ions that requires $\kappa_{i,e}\leftrightarrow\kappa_{e,i}$
and $\vec{U}_{0}\leftrightarrow -\vec{U}_{0}$. Equation
(\ref{Omega_K}) for the Doppler-shifted frequency in the ion
frame, $\Omega_{\vec{k}}^{(i)}$, is symmetric with respect to a
similar expression
for the wave frequency in the electron frame, $\Omega_{\vec{k}}^{(e)}%
\equiv\omega_{\vec{k}}-\vec{k}\cdot\vec{V}_{e0}=\Omega_{\vec{k}}^{(i)}%
-\vec{k}\cdot\vec{U}_{0}$,
\begin{equation}
\Omega_{\vec{k}}^{(e)}=-\ \frac{(1+\kappa_{i}^{2})[1+(1+\kappa_{e}%
^{2})k_{\parallel}^{2}/k_{\perp}^{2}](\vec{k}\cdot\vec{U}_{0})}{\kappa
_{i}\left(  \kappa_{i}+\kappa_{e}\right)  (1+\psi_{\vec{k}})}.
\label{Omega_k-kV_i0}%
\end{equation}
Notice that for prevalent waves with positive
$\vec{k}\cdot\vec{U}_{0}$ the shifted electron frequency
$\Omega_{\vec{k}}^{(e)}$ is always negative, while the
corresponding ion frequency $\Omega_{\vec{k}}^{(i)}$ is
positive. This reflects the fact that such waves, regardless of
the particle magnetization, always lag behind the streaming
electrons but move ahead of the ions.

Despite the formal symmetry in the general relations between
electrons and ions, their actual contributions are not
equivalent. In the lower ionosphere, since $m_{i}\simeq30$\,amu
and $\nu_{e}/\nu_{i}\simeq10$ \citep{Kelley:Ionosphere2009},
the ratio $\kappa_{e}/\kappa_{i}$ is huge,
$\kappa_{e}/\kappa_{i}\simeq5500$. Then practically everywhere,
except for the \emph{D}-region altitudes below 80~km, electrons
are strongly magnetized, $\kappa_{e}\gg1$, whilst in most of
the \emph{E}/\emph{D}-region electrojet ions are largely
unmagnetized, $\kappa _{i}\ll1$, reaching only a partial
magnetization, $\kappa_{i}\gtrsim1$, above 110~km. Apart from
the \emph{D}-region altitudes but practically throughout the
entire electrojet, setting $\kappa_{e}\gg1$ with arbitrary
$\kappa_{i}$, we reduce Eqs.~(\ref{psi}) to (\ref{omega_k}) to
simpler relations \citep{Fejer:Theory84}:
\begin{align}
\psi_{\vec{k}}  &  \approx\psi_{\perp}\left[  1+(1+\kappa_{i}^{2})\frac
{\Omega_{e}^{2}k_{\parallel}^{2}}{\nu_{e}^{2}k_{\perp}^{2}}\right]
,\nonumber\\
\omega_{\vec{k}}  &  \approx\frac{\vec{k}\cdot\vec{U}_{0}}{1+\psi_{\vec{k}}%
}-\kappa_{i}\vec{k}\cdot(\vec{U}_{0}\times\hat{b}),\label{for_kappa_e>>1}\\
\delta\Phi_{\vec{k},\omega}  &  =i\ \frac{m_{i}\nu_{i}(1+\kappa_{i}^{2}%
)(\vec{k}\cdot\vec{U}_{0})}{ek_{\perp}^{2}(1+\psi_{\vec{k}})}\left(
\frac{\delta n_{\vec{k},\omega}}{n_{0}}\right)  .\nonumber
\end{align}
For unmagnetized ions, $\kappa_{i}\ll1$, Eq.~(\ref{for_kappa_e>>1}) reduces
further to the conventional relations \citep{Farley:96}.

Recall that all these relations represent only the first-order
approximation with respect to the small parameters $kl_{i}$ and
$\omega/\nu_{i}$. Neglect of pressure gradients, particle
inertia, etc., results in the highest-order dispersion relation
for the real part of the linear wave frequency,
Eq.~(\ref{omega_k}). It contains no $\pi/2$-phase shifted
corrections that determine the linear wave growth or damping.
For arbitrary particle magnetization, the corresponding
second-order corrections to the dispersion relation for the FB
and gradient drift instabilities are obtained in the appendix
of \citet{Dimant:Magnetosphere2011}. We should bear in mind,
however, that our quasi-periodic description of the
non-linearly saturated steady state requires all wave
frequencies to be strictly real. This implies that the
second-order destabilizing linear factors are on average
balanced by non-linearities. Furthermore, non-linearly
saturated turbulence itself has a strongly time-varying
character caused by dynamic mode-coupling
\citep[e.g.,][]{Hamza:turbulent93,Hamza:self93,Dimant:2000}.
All this breaks the narrowband approximation for which any wave
frequency $\omega$, via the linear dispersion relation, is
uniquely determined by the corresponding wavevector $\vec{k}$,
$\omega\approx\omega_{\vec{k}}$. This means that the
$\delta$-function dependence of the frequency spectrum is
actually spread over a finite band around $\omega _{\vec{k}}$.
Accurate theoretical description of such non-linearly saturated
states is extremely difficult, especially for strong turbulence
generated by the driving field well above the instability
threshold. To avoid serious difficulties and mathematical
complexities, we will continue using first-order linear
relations like Eqs.~(\ref{delta_Vs}) and
(\ref{Phi->del_n_symmetric}) and neglect next-order linear and
non-linear corrections. We expect the corresponding errors to
be reasonably small, although these expectations need
additional testing.

\subsection{Non-linear Currents\label{non-linear current}}

As shown in Sect.~\ref{General consideration}, the non-linear
current (NC) plays an important role in energy conversion. To
further clarify its physical meaning, we calculate separately
NCs for electrons and ions, $\vec{j}_{s}^{\mathrm{NC}}$
($s=e,i$), in terms of a given density-irregularity spectrum,
$\delta n_{\vec{k},\omega}$. These partial currents will be
used in Sect.~\ref{Energy inputs and turbulent heating} for
comparison of the partial turbulent Joule heating,
$\vec{E}_0\cdot\vec{j}_{s}^{\mathrm{NC}}$, with the
corresponding frictional heating terms.

Using Eqs.~(\ref{Average_harmonics}), (\ref{delta_Vs}), and
(\ref{Phi->del_n_symmetric}), we obtain the partial electron NC
density, $\vec {j}_{e}^{\mathrm{NC}}\equiv-e\langle\delta
n\delta\vec{V}_{e}\rangle =-e\sum_{\vec{k},\omega\neq0}\delta
n_{\vec{k}}^{\ast}\delta\vec{V}_{e\vec{k}}$:
\begin{align}
\vec{j}_{e}^{\mathrm{NC}} &  =\frac{(1+\kappa_{e}^{2})\left(  1+\kappa_{i}%
^{2}\right)  en_{0}}{\kappa_{i}\left(  \kappa_{e}+\kappa_{i}\right)  }%
\sum_{\vec{k},\omega\neq0}\frac{\vec{k}\cdot\vec{U}_{0}}{(1+\psi_{\vec{k}%
})k_{\perp}^{2}}\nonumber\\
&  \times\left[  \vec{k}_{\parallel}+\frac{\vec{k}_{\perp}-\kappa_{e}(\vec
{k}_{\perp}\times\hat{b})}{1+\kappa_{e}^{2}}\right]  \left\vert \frac{\delta
n_{\vec{k},\omega}}{n_{0}}\right\vert ^{2}.\label{j_NLe}%
\end{align}
Similarly, we obtain the partial ion NC density, $\vec{j}_{i}^{\mathrm{NC}}\equiv
e\langle\delta n\delta\vec{V}_{i}\rangle\allowbreak=\allowbreak e\sum_{\vec
{k},\omega\neq0}\delta n_{\vec{k}}^{\ast}\delta\vec{V}_{i\vec{k}}$, which
differs from Eq.~(\ref{j_NLe})
by the replacement $\kappa_{e,i}\leftrightarrow\kappa_{i,e}$ and
the ``plus'' sign in front of $(\vec{k}_{\perp}\times\hat{b})$.
The total NC, $\vec{j}^{\mathrm{NC}}\equiv-e\langle\delta n\delta\vec
{U}\rangle=-e\sum_{\vec{k},\omega\neq0}\delta n_{\vec{k}}^{\ast}\delta\vec
{U}_{\vec{k}}$, is then given by
\begin{align}
&  \vec{j}^{\mathrm{NC}}=\vec{j}_{e}^{\mathrm{NC}}+\vec{j}_{i}^{\mathrm{NC}%
}=\frac{en_{0}}{\kappa_{i}\kappa_{e}}\sum_{\vec{k},\omega\neq0}
(\vec{k}\cdot\vec{U}_{0})\left\vert \frac{\delta n_{\vec{k},\omega}}{n_{0}}
\right\vert^{2}
\nonumber\\
&\times  \frac{(1+\kappa_{e}^{2})\left(
1+\kappa_{i}^{2}\right)  \vec{k}_{\parallel}+\left(  1+\kappa_{i}\kappa
_{e}\right)  \vec{k}_{\perp}-\left(  \kappa_{e}-\kappa_{i}\right)  (\vec
{k}\times\hat{b})}{(1+\psi_{\vec{k}%
})k_{\perp}^{2}}. 
\label{j_NL}%
\end{align}
At altitudes above $100$~km, where both
$\kappa_{e}\gg1\gtrsim\kappa_{i}$ and
$\kappa_{i}\kappa_{e}=\psi_{\perp}^{-1}\gg1$ hold together,
according to Eq.~(\ref{U_0}), we have
\begin{equation}
\vec{U}_{0}=\frac{\vec{E}_{0}\times\hat{b}-\kappa_{i}\vec{E}_{0}}{\left(
1+\kappa_{i}^{2}\right)  B},\label{U_0_simple}%
\end{equation}
while the ion non-linear current turns out to be negligible compared to that of
electrons, $\vec{j}_{e}^{\mathrm{NC}}\approx\vec{j}^{\mathrm{NC}}$. We expect
no spectral asymmetry along $\vec{B}$, so that $\sum_{\vec{k},\omega\neq
0}f(k_{\perp}^{2})\vec{k}_{\parallel}=0$. As a result, the
NC density, $\vec{j}^{\mathrm{NC}}\perp\vec{B}$, reduces in this limit to
\begin{equation}
\vec{j}^{\mathrm{NC}}\approx-\ \frac{en_{0}}{\kappa_{i}}\sum_{\vec{k}%
,\omega\neq0}\frac{(\vec{k}\times\hat{b})(\vec{k}\cdot\vec{U}_{0})}%
{(1+\psi_{\vec{k}})k_{\perp}^{2}}\left\vert \frac{\delta n_{\vec{k},\omega}%
}{n_{0}}\right\vert ^{2}.\label{non-linear_simplest}%
\end{equation}
Within the bulk electrojet where ions are unmagnetized, $\kappa_{i}\ll1$, the
$\vec{k}$-spectrum of irregularities is largely perpendicular to $\vec{E}_{0}%
$, so that $\vec{j}^{\mathrm{NC}}$ has there a predominantly
Pedersen direction. This is of paramount importance for the
global MI coupling \citep{Dimant:Magnetosphere2011}.

\subsection{Partial Energy Inputs and Turbulent
Heating\label{Energy inputs and turbulent heating}}

Calculating the partial energy inputs and frictional heating
sources for a specific plasma model and given turbulence
spectrum allows to quantitatively understand how the turbulent
energy is distributed between different plasma components. Here
we obtain such expressions for arbitrarily magnetized two-fluid
plasmas in the quasilinear approximation and verify, in
particular, that exact Eqs.~(\ref{so_that_really}) and
(\ref{L_turb}) remain exactly valid. The turbulent frictional
heating sources found here could  be included into
ionosphere-thermosphere computer models, as explained in the
companion paper \citep{Dimant:Magnetosphere2011}.

Multiplying Eq.~(\ref{Lorentz}) by the plasma density and
corresponding fluid velocities gives
\begin{equation}
\vec{E}\cdot\vec{j}_{s}=m_{s}\nu_{s}nV_{s}^{2}
\label{Heating}%
\end{equation}
($s=e,i$), where $\vec{j}_{e}\equiv-en\vec{V}_{e}$ and
$\vec{j}_{i}\equiv en\vec{V}_{i}$ are the electron and ion
partial current densities. These expressions relate the work
done by the field $\vec{E}$ on the \textit{s}-particle currents
to the corresponding sources of frictional heating. As
mentioned in Sect.~\ref{General consideration}, the actual
frictional heating of $s$-type particles is smaller than
$m_{s}\nu_{s n}nV_{s}^{2}$ because a fraction of the acquired
field energy equal to $m_s/(m_s + m_n)$ goes immediately to the
colliding neutrals without heating the plasma particles
\citep{Schunk:Ionospheres09}. This is especially true for ions
with $m_i\simeq m_n$, whose frictional heating source is nearly
half of $m_{i}\nu_{i}nV_{i}^{2}$. In what follows, we will
refer to $m_{s}\nu_{s n}nV_{s}^{2}$ as the \textit{s-n}
`heating term' with the caveat that the actual $s$-particle
frictional heating is described by $m_n/(m_s + m_n)$ of
$m_{s}\nu_{s}nV_{s}^{2}$, while the remaining fraction goes to
neutral ($n$) frictional heating
\citep{Dimant:Magnetosphere2011}.

Equation~(\ref{Heating}) includes total field energy losses and
plasma heating. The zero-order heating alone, i.e., that without any
plasma turbulence, is described by
\begin{equation}
\vec{E}_{0}\cdot\vec{j}_{s0}=m_{s}\nu_{s}n_{0}V_{s0}^{2}.
\label{Heating_0}%
\end{equation}
Subtracting Eq.~(\ref{Heating_0}) from Eq.~(\ref{Heating}) and averaging yields
the turbulent heating rates,
\begin{eqnarray}
\vec{E}_{0}\cdot\vec{j}_{s}^{\mathrm{NC}}
\!\!\!& + &\!\!\!\langle\delta\vec{E}\cdot\delta\vec{j}_{s}\rangle\nonumber
\\
& = &\!\! m_{s}\nu_{s}(n_{0}\langle \delta V_{s}^{2}\rangle +2\vec
{V}_{s0}\cdot\langle\delta\vec{V}_{s}\delta n\rangle+\!\langle \delta
V_{s}^{2}\delta n\rangle),%
\label{HH}%
\end{eqnarray}
where $\vec{j}_{e}^{\mathrm{NC}}=-e\langle\delta
n\delta\vec{V}_{e}\rangle$ and
$\vec{j}_{i}^{\mathrm{NC}}=e\langle\delta
n\delta\vec{V}_{i}\rangle$ are the partial electron and ion
contributions to the total NC given by Eq.~(\ref{j_NL}). The
LHS of Eq.~(\ref{HH}) describes the additional average work by
electric fields in the turbulent plasma on a given plasma
species. The RHS represents the corresponding average turbulent
heating per unit volume. It is not obvious, however, that the
approximate quasilinear expressions obtained above ensure that
the two sides of Eq.~(\ref{HH}) are really equal. To verify the
equality and clarify the physical meaning of various terms, we
will calculate the two sides of Eq.~(\ref{HH}) separately.

We start by calculating the RHS of Eq.~(\ref{HH}) that
describes the frictional heating. The positively determined and
dominant term $m_s\nu_sn_{0}\langle\delta V_{s}^{2}\rangle$
describes energization of individual plasma particles, while
the other terms are associated with density variations at a
given location. Consistency of the quasilinear approximation
requires neglecting the last, cubically non-linear, term. To
calculate average quadratically non-linear quantities in terms
of given irregularity spectra, we  use
Eq.~(\ref{Average_harmonics}).

By using Eq.~(\ref{delta_Vs}), for the dominant electron
and ion heating terms, we obtain
\begin{equation}
m_{s}\nu_{s}n_{0}\left\langle \delta V_{s}^{2}\right\rangle   =\frac
{en_{0}\kappa_{s}}{B}\sum_{\vec{k},\omega\neq0}\left(  \frac{k_{\perp}^{2}%
}{1+\kappa_{s}^{2}}+k_{\parallel}^{2}\right)  |\delta\Phi_{\vec{k},\omega
}|^{2}. %
\label{intermediate}%
\end{equation}
If $\kappa_{e}\gg\kappa_{i}$ and $\kappa_{e}\gg1$ then the ion
perpendicular heating dominates over the corresponding electron
one, while for the parallel heating the reverse is true. At
lower \emph{D}-region altitudes, where electrons are partially
demagnetized, $\kappa_{i}\ll\kappa_{e}\lesssim1$, the electron
heating always prevails. Using
Eq.~(\ref{Phi->del_n_symmetric}), we rewrite
Eq.~(\ref{intermediate}) in terms of a given
density-irregularity spectrum as
\begin{align}
m_{e}\nu_{e}n_{0}\left\langle \delta V_{e}^{2}\right\rangle  &  =\frac
{en_{0}B(1+\kappa_{e}^{2})(1+\kappa_{i}^{2})^{2}}{(\kappa_{e}+\kappa_{i}%
)^{2}\kappa_{i}^{2}\kappa_{e}}\nonumber\\
&  \times\sum_{\vec{k},\omega\neq0}\frac{[1+(1+\kappa_{e}^{2})k_{\parallel
}^{2}/k_{\perp}^{2}](\vec{k}\cdot\vec{U}_{0})^{2}}{(1+\psi_{\vec{k}}%
)^{2}k_{\perp}^{2}}\left\vert \frac{\delta n_{\vec{k},\omega}}{n_{0}%
}\right\vert ^{2}
\label{diag_psi}%
\end{align}
and similar for ions with the symmetric replacement
$\kappa_{e,i}\leftrightarrow\kappa_{i,e}$. Using
Eqs.~(\ref{Phi->del_n_symmetric}) and (\ref{used_here}), for
the combined heating rate we obtain
\begin{subequations}
\label{diag_psi_summa_ei}
\begin{align}
&  m_{e}\nu_{e}n_{0}\left\langle \delta V_{e}^{2}\right\rangle +m_{i}\nu
_{i}n_{0}\left\langle \delta V_{i}^{2}\right\rangle \nonumber\\
&  =\frac{eBn_{0}\psi_{\perp}(1+\kappa_{e}^{2})(1+\kappa_{i}^{2})}{\kappa
_{e}+\kappa_{i}}\sum_{\vec{k},\omega\neq0}\frac{(\vec{k}\cdot\vec{U}_{0})^{2}%
}{(1+\psi_{\vec{k}})k_{\perp}^{2}}\left\vert \frac{\delta n_{\vec{k},\omega}%
}{n_{0}}\right\vert ^{2}\label{diag_psi_summa_ei_1}\\
&  =\frac{en_{0}\left(  \kappa_{i}+\kappa_{e}\right)  }{B\psi_{\perp}\left(
1+\kappa_{e}^{2}\right)  (1+\kappa_{i}^{2})}\sum_{\vec{k},\omega\neq0}(1+\psi
_{\vec{k}})k_{\perp}^{2}\left\vert \delta\Phi_{\vec{k},\omega}\right\vert
^{2}.\label{diag_psi_summa_ei_2}
\end{align}
\end{subequations}

Equation~(\ref{diag_psi_summa_ei}), presented here in terms of both $\delta
n_{\vec{k},\omega}$ and $\delta\Phi_{\vec{k},\omega}$, allows
one to understand the quantitative difference between 2-D and
3-D turbulent heating. The 3-D effect of parallel-field dominated turbulent heating
\citep{DimantMilikh:JGR03,MilikhDimant:JGR03} is entirely due
to the difference between $\psi_{\vec{k}}$ and $\psi_{\perp}$
in the multipliers $(1+\psi_{\vec{k}})$. While the total
perpendicular turbulent heating is
described by the $(1+\psi_{\perp})$ component, the total parallel turbulent
heating derives from the remainder, $\psi_{\vec{k}}%
-\psi_{\perp}=\psi_{\perp}(1+\kappa_{e}^{2})(1+\kappa_{i}^{2})k_{\parallel
}^{2}/k_{\perp}^{2}$, as seen from Eq.~(\ref{psi}). If $|\delta
n_{\vec{k},\omega}|^{2}$ were equal in 2-D and 3-D then,
according to Eq.~(\ref{diag_psi_summa_ei_1}), larger $(1+\psi
_{\vec{k}})$ would reduce total turbulent heating in 3-D
compared to 2-D. We have, however, the reverse by the following
reason. Heuristic arguments require the perpendicular rms
turbulent fields $\sim\sum_{\vec{k},\omega\neq0}
k_{\perp}^{2}|\delta\Phi_{\vec{k},\omega}|^{2}$ to be
approximately equal in 2-D and 3-D
\citep{DimantMilikh:JGR03,Dimant:Magnetosphere2011}. According
to Eq.~(\ref{diag_psi_summa_ei_2}), this leads to stronger
heating in 3-D compared to 2-D, entirely due to larger density
perturbations, $\sum_{\vec{k},\omega\neq0} |\delta
n_{\vec{k},\omega}|^{2}\propto\sum_{\vec{k},\omega\neq0}
(1+\psi_{\vec{k}})^{2}k_{\perp}^{2}|
\delta\Phi_{\vec{k},\omega}|^{2}$. The same pertains to the
non-linear currents discussed in Sect.~\ref{non-linear
current}. All these heuristic inferences have been confirmed by
our recent supercomputer PIC simulations
\citep{Oppenheim:Fully2011}.

The second term in the RHS of Eq.~(\ref{HH}), calculated
by expressing $\delta\vec{V}_{e,i}$ and $\vec{V}_{e,i0}$ in terms
of $\delta \vec{U}$ and $\vec{U}_{0}$, using Eq.~(\ref{V_perp_via_U}),
and $\vec{U}_{0}\times\hat{b}$ in terms of $\vec{E}_{0}$, using
Eq.~(\ref{ExB_via_U}), is
\begin{eqnarray}
2m_{i}&\!\!\!\!\!\!(&\!\!\!\!\!\!\nu_{i}\vec{V}_{i0}\cdot \langle\delta
n\delta\vec{V}_{i}\rangle\,)\nonumber\\
& \!= &\!\frac{2en_{0}(1+\kappa_{e}^{2})}{\kappa_{e}(\kappa_{i}+\kappa_{e})}
\sum
_{\vec{k},\omega\neq0}\frac{(\vec{k}\cdot\vec{E}_{0})(\vec{k}\cdot\vec{U}
_{0})}{(1+\psi_{\vec{k}})k_{\perp}^{2}}\left\vert \frac{\delta
n_{\vec {k},\omega}}{n_{0}}\right\vert ^{2}%
\label{``diag''_2}
\end{eqnarray}
and similar for electrons, with the symmetric replacement
$\kappa_{e,i}\leftrightarrow\kappa_{i,e}$. At higher altitudes,
where $\psi_{\perp}=(\kappa_{e}\kappa_{i})^{-1}<1$ and
$\kappa_{e}\gg1$, the ion component dominates over the electron
component, while at lower altitudes, $\psi_{\perp}>1$, the
reverse holds. Adding up the two plasma components gives
\begin{align}
&2m_{i}\nu_{i}(\vec{V}_{i0}\cdot\langle\delta n\delta\vec{V}_{i}\rangle) +
 2m_{e}\nu_{e}(\vec{V}_{e0}\cdot\langle\delta n\delta\vec{V}_{e}
\rangle)\nonumber\\
&  =2en_{0}\left(  1+\psi_{\perp}\right)  \sum_{\vec{k},\omega\neq0}
\frac{(\vec{k}\cdot\vec{E}_{0})(\vec{k}\cdot\vec{U}_{0})}{(1+\psi_{\vec{k}
})k_{\perp}^{2}}\left\vert \frac{\delta n_{\vec{k},\omega}}{n_{0}}\right\vert
^{2}. \label{``diag''_combined}%
\end{align}
In the most of electrojet, these terms are small compared to
those in Eq.~(\ref{diag_psi_summa_ei}), although at the top of
the electrojet they can become comparable.

To verify that general Eqs.~(\ref{so_that_really}) and
(\ref{L_turb}) exactly hold in our quasilinear calculations, now we
proceed to calculating the LHS of Eq.~(\ref{HH}). Each of the
electron and ion terms,
$\langle\delta\vec{E}\cdot\delta\vec{j}_{e,i}\rangle$, can be
separated into two distinct parts,
\begin{equation}
\langle\delta\vec{E}\cdot\delta\vec{j}_{s}\rangle=\langle\delta\vec{E}
\cdot\delta\vec{j}_{s}\rangle_{1}+\langle\delta
\vec{E}\cdot\delta\vec{j}_{s}\rangle_{2},%
\label{separa}
\end{equation}
where $\langle\delta\vec{E}\cdot\delta\vec{j}_{s}\rangle_{1}
\equiv-en_{0}
\langle\delta\vec{V}_{s}\cdot\delta\vec{E}\rangle$, $
\langle\delta\vec{E}\cdot\delta\vec{j}_{s}\rangle_{2}
\equiv-e\vec{V} _{e0}\cdot\langle\delta n\delta\vec{E}\rangle$,
and we neglect the cubically non-linear corrections,
$e\langle\delta n\delta\vec{V}_{s}\cdot\delta\vec{E}\rangle$.
Using Eqs.~(\ref{ExB_via_U}) and (\ref{delta_Vs}), we find that
the first term,
$\langle\delta\vec{E}\cdot\delta\vec{j}_{s}\rangle_{1}$, equals
the fixed-density frictional heating rate,
\begin{equation}
\langle\delta\vec{E}\cdot\delta\vec{j}_{s}\rangle_{1}=m_{s}\nu_{s}%
n_{0}\left\langle \delta V_{s}^{2}\right\rangle, \label{ravny}%
\end{equation}
while the combination of the electron and ion second terms in
Eq.~(\ref{separa}) yields
\begin{align}
&  \langle\delta\vec{E}\cdot\delta\vec{j}_{e}\rangle_{2}+\langle\delta\vec{E}
\cdot\delta\vec
{j}_{i}\rangle_{2} \equiv\langle\delta\vec{E}\cdot\delta\vec{j}\rangle_{2}
\nonumber\\
&  =-\ \frac{eBn_{0}\left(  1+\kappa_{e}^{2}\right)  \left(  1+\kappa_{i}%
^{2}\right)  \psi_{\perp}}{\left(  \kappa_{e}+\kappa_{i}\right)  }\sum
_{\vec{k},\omega\neq0}\frac{(\vec{k}\cdot\vec{U}_{0})^{2}}{(1+\psi_{\vec{k}}%
)k_{\perp}^{2}}\left\vert \frac{\delta n_{\vec{k}}}{n_{0}}\right\vert ^{2}
\label{del_22}%
\end{align}
By combining Eqs.~(\ref{diag_psi_summa_ei}) and
(\ref{separa})--(\ref{del_22}), we verify that, in accord with
Eq.~(\ref{so_that_really}),
$\langle\delta\vec{E}\cdot\delta\vec{j}\rangle=0$.
Due to this, turbulent Joule heating, $L_{\mathrm{turb}}\equiv
\langle \vec{E}\cdot\vec{j}\rangle - \vec{E}_0\cdot\vec{j}_0 =
\vec{E}_0\cdot\vec{j}^{\mathrm{NC}} +
\langle\delta\vec{E}\cdot\delta\vec{j}\rangle$, in accord with
Eq.~(\ref{L_turb}), equals the total average work of the
external electric field on the total non-linear current,
$P_{\mathrm{NC}}\equiv\vec{E}_0\cdot\vec{j}^{\mathrm{NC}}$. This
quantity, in turn, should equal the total frictional heating
source given by combining the RHS of Eq.~(\ref{HH}) for all
plasma species. Indeed, using Eq.~(\ref{j_NL}) for
$\vec{j}^{\mathrm{NC}}$, we obtain
\begin{align}
P_{\mathrm{NC}}  &
=\frac{eBn_{0}
\psi_{\perp}(1+\kappa_{e}^{2})(1+\kappa_{i}^{2})}{\kappa_{e}+\kappa_{i}}%
\sum_{\vec{k},\omega\neq0}\frac{(\vec{k}\cdot\vec{U}_{0})^{2}}{(1+\psi_{\vec{k}%
})k_{\perp}^{2}}\left\vert \frac{\delta n_{\vec{k}}}{n_{0}}\right\vert
^{2}\nonumber\\
&  +2(1+\psi_{\perp})en_{0}\sum_{\vec{k},\omega\neq0}\frac{(\vec{k}\cdot\vec{E}%
_{0})(\vec{k}\cdot\vec{U}_{0})}{(1+\psi_{\vec{k}})k_{\perp}^{2}}\left\vert
\frac{\delta n_{\vec{k}}}{n_{0}}\right\vert ^{2}, \label{total_P_NL_again}%
\end{align}
which equals the total turbulent heating found by combining
Eqs.~(\ref{diag_psi_summa_ei}) and (\ref{``diag''_combined}).

Thus, our fluid-model calculations for arbitrarily magnetized
plasma particles in the quasilinear approximation have fully
confirmed the exact general relations derived in
Sect.~\ref{General consideration} exclusively from Maxwell's
equations. The partial relations for various energy conversion
terms show how the deposited field energy is divided between
the electron and ion heating channels and allow one to properly
interpret the corresponding terms.

\section{Global Energy Flow\label{Global Energy Flow}}

Now we discuss how the energy from the Earth's magnetosphere is deposited to
the \emph{E}/\emph{D}-region electrojet. From the above treatment it is clear
that the global energy flow should be analyzed based on the total average
currents that include the NCs. To obtain energy
balance equations similar to Eqs.~(\ref{Poynting's}) for spatially and
temporally averaged fields introduced in Eq.~(\ref{perturbations}), we use
separate linear Maxwell's equations for the average fields and for disturbances
caused by the total average currents, $\langle\vec{j}\rangle=\vec
{j}^{\mathrm{tot}}\equiv\vec{j}_{0}+\vec{j}_{\mathrm{NL}}$, as we did in
deriving Eq.~(\ref{Poynting's_perturb}) for the wave perturbations. As a
result, ignoring the induction component of $\langle\vec{E}\rangle=\vec{E}%
_{0}$, i.e., assuming $\nabla_{\vec{r}}\times\vec{E}_{0}=0$, we obtain
\begin{equation}
\partial_{\tau}\left[  \frac{\varepsilon_{0}\left(  E_{0}^{2}+c^{2}\langle
B\rangle^{2}\right)  }{2}\right]  +\nabla_{\vec{r}}\cdot\Delta\vec{S}=-\vec
{E}_{0}\cdot\vec{j}^{\mathrm{tot}}, \label{field_energy_transport}%
\end{equation}
where $\tau$, $\vec{r}$ are the large-scale variables introduced in
Sect.~\ref{General consideration}, and
\begin{equation}
\Delta\vec{S}\equiv\frac{\vec{E}_{0}\times\Delta\vec{B}}{\mu_{0}}=\epsilon
_{0}c^{2}\vec{E}_{0}\times\Delta\vec{B}. \label{Delta_S}%
\end{equation}
Here
$\Delta\vec{B}\equiv\langle\vec{B}\rangle-\vec{B}_{0}$ is the quasi-stationary large-scale magnetic field disturbance, where by $\vec{B}_{0}$
we mean the geomagnetic field undisturbed by the electrojets, $\nabla_{\vec
{r}}\times\vec{B}_{0}=0$, so that $\nabla_{\vec{r}}\cdot\left(  \vec{E}%
_{0}\times\vec{B}_{0}\right)  =0$. The disturbance
$\Delta\vec{B}$ is caused by the total average electrojet
currents,
$\nabla\times\Delta\vec{B}\approx\mu_{0}\vec{j}^{\mathrm{tot}}$.
It is usually so small, $|\Delta\vec{B}|\lesssim10^{-3}B_{0}$,
that can be neglected in the expression for the magnetic energy
density, $\langle B\rangle^{2}\approx B_{0}^{2}$. However,
$\Delta\vec{B}$ is crucial for the Poynting flux
$\Delta\vec{S}$ that provides a downward flow of field energy
from the magnetosphere to the \emph{E}/\emph{D}-region
ionosphere. Further, bearing in
mind that $\vec{j}^{\mathrm{tot}}=\overleftrightarrow{\sigma^{\mathrm{tot}}%
}\cdot\vec{E}_{0}$, the work of the external field on the total current in the
RHS\ of Eq.~(\ref{field_energy_transport}) can be written as $\vec{E}_{0}%
\cdot\vec{j}^{\mathrm{tot}}=\sigma_{\mathrm{P}}^{\mathrm{tot}}E_{0}^{2}$, where $\overleftrightarrow{\sigma^{\mathrm{tot}}}$ and $\sigma_{\mathrm{P}}^{\mathrm{tot}}$ are the total conductivity tensor and Pedersen conductivity, respectively \citep{Dimant:Magnetosphere2011}. In the turbulent electrojet, the quantity $\sigma_{\mathrm{P}}^{\mathrm{tot}}E_{0}^{2}$ represents the combined laminar and turbulent Joule heating. Bearing in mind quasi-stationary
conditions, from Eq.~(\ref{field_energy_transport}) we estimate the magnitude
of the Poynting flux on top of the electrojet as $|\Delta\vec{S}|\sim
\sigma_{\mathrm{P}}^{\mathrm{tot}}E_{0}^{2}L_{\parallel}$, where
$L_{\parallel}$ is the characteristic size of the electrojet along the nearly
vertical magnetic field. The corresponding disturbance of the magnetic field,
$\Delta\vec{B}$, has a significant component perpendicular to both $\vec
{E}_{0}$ and $\vec{B}_{0}$ with the magnitude $\sim\mu_{0}\sigma_{\mathrm{P}%
}^{\mathrm{tot}}E_{0}L_{\parallel}$.

Now we do the same for the plasma. Averaging
Eq.~(\ref{particle_energy _conserva}) over small turbulent
scales and neglecting
$\langle\delta\vec{E}\cdot\delta\vec{j}\rangle$, in accord with
the analysis of Sect.~\ref{General consideration}, we obtain
\begin{equation}
\partial_{\tau}\sum_{s}\langle\mathcal{E}_{s}\rangle+\nabla_{r}%
\cdot\sum_{s}\langle\vec{K}_{s}\rangle=\vec{E}_{0}\cdot\vec
{j}^{\mathrm{tot}}+\sum_{s}\langle L_{s}\rangle.
\label{particle_energy_transport}%
\end{equation}
While the RHS of Eq.~(\ref{field_energy_transport}) has only one term
describing the total field energy loss to charged particles, the RHS\ of
Eq.~(\ref{particle_energy_transport}) has two terms. The first term, $\vec{E}_{0}\cdot\vec{j}^{\mathrm{tot}}=\sigma_{\mathrm{P}}%
^{\mathrm{tot}}E_{0}^{2}$, describes the total Joule heating
rate due to the total energy input from the fields. The second term describes
particle energy dissipation in the abundant neutral atmosphere. For
particle fluxes dominated by the thermal bulk, the two terms almost
cancel each other locally, so that the flux divergence term $\nabla_{r}%
\cdot\sum_{s}\langle\vec{K}_{s}\rangle$ is expected to be much
less than $\nabla_{\vec{r}}\cdot\Delta\vec{S}$ in
Eq.~(\ref{field_energy_transport}). This means that the energy
transport of ionospheric particles should be much weaker than
that of fields. The reason why the particle energy transport
plays a minor role is associated with relatively short mean
electron and ion free paths, even in the almost vertical
direction along $\vec{B}_{0}$. We should bear in mind, however,
that the fluxes $\langle\vec{K}_{s}\rangle$ can also include
high-energy precipitating particles that may comprise a
significant part of the field-aligned (Birkeland) currents
within auroral arcs. These occasions may break the nearly
perfect local balance between the Joule heating and atmospheric
cooling. Although in these cases $|\nabla_{r}\cdot\sum_{s
}\langle\vec{K}_{s}\rangle|$ should still remain less than
$|\nabla _{\vec{r}}\cdot\Delta\vec{S}|$, the two fluxes may be
closer to each other than those for the nearly local
thermal-bulk dominated balance.

Now we compare the total field and particle fluxes along the nearly vertical
magnetic field in greater detail. We can estimate the particle energy flux
magnitude as $| \sum_{s}\vec{K}_{s}| \sim
\langle\mathcal{E\rangle}j_{\parallel}$, where $\langle\mathcal{E\rangle}$ is
an effective energy of charged particles. Using the charge flow conservation,
$\nabla\cdot\vec{j}=0$, we estimate the parallel current density as
$j_{\parallel}\sim L_{\parallel}\sigma_{P}E_{0}/L_{\mathrm{P}}$, where
$L_{\mathrm{P}}$ is the characteristic scale of current density variations in
the Pedersen direction. As a result, we have $|\sum_{s}\langle\vec
{K}_{s}\mathcal{\rangle}|/|\Delta\vec{S}|\sim\langle\mathcal{E\rangle
}/(e|\Delta\Phi|)$, where $\Delta\Phi\sim L_{\mathrm{P}}E_{0}$ is the
characteristic cross-polar cap potential. Typical values of $\Delta\Phi$ are
$\sim100$~kV. If this current is mainly provided by thermal bulk particles
then their energy is many orders of magnitude smaller, $\langle
\mathcal{E\rangle}\lesssim1$~eV. Precipitating particles can have a rather
high energy, $\langle\mathcal{E\rangle}\lesssim30$~keV \citep[e.g.,][]{Ashrafi2005:Comparison}, so that the particle energy flux can sometimes be comparable to the Poynting flux.

\citet{Buchert:Effect2006} in their analysis of global energy flow between the
magnetosphere and \emph{E}-region ionosphere put more stress on the Birkeland
currents. These field-aligned currents are important for charge conservation
and they provide the MI coupling via particle
precipitation. Furthermore, electron fluxes along $\vec{B}_{0}$, because of
their high parallel mobility, ensure effective mapping of the electric field
from the magnetosphere to lower ionosphere. From the energy transport
viewpoint, however, our analysis gives preference to the Poynting flux.
It is interesting to note that 2-D or 3-D simulations of the \emph{E}%
/\emph{D}-region instabilities in purely periodic boxes have
always generated the DC currents $\vec{j}^{\mathrm{tot}}$ in a
plane perpendicular to $\vec {B}_{0}$. According to Maxwell's
equations, $\vec{j}^{\mathrm{tot}}$ must generate
quasi-stationary loop-like magnetic disturbances
$\Delta\vec{B}$, which in the employed electrostatic codes are
ignored by definition. Furthermore, such $\Delta\vec{B}$ would
even violate the imposed periodicity, unless one presumes on
the 3-D box boundaries surface currents that exactly balance
the volumetric currents. If, however, one did not ignore
$\Delta\vec{B}$ then the corresponding Poynting flux,
$c^{2}\epsilon_{0}\vec {E}_{0}\times\Delta\vec{B}$, would be
directed inward the box and provide the required energy input.

\section{Summary and Conclusions}

Plasma turbulence generated by \emph{E}/\emph{D}
-region instabilities can contribute significantly to the global energy
exchange between the magnetosphere and ionosphere. The spatially-temporally averaged energy deposit in the turbulent electrojet is
given by $\vec{E}_{0}\cdot\vec{j}^{\mathrm{tot}}$, where the total current
density $\vec{j}^{\mathrm{tot}}$, in addition to the regular current density, $\vec
{j}_{0}=\overleftrightarrow{\sigma_{\mathrm{P}}^{0}}\cdot\vec{E}_{0}$, includes
the non-linear current (NC) density, $\vec{j}^{\mathrm{NC}}$, caused by low-altitude plasma turbulence \citep{RogisterJamin:1975,Oppenheim:Evidence97}. The work of the external field on the NC, $\vec{E}_{0}\cdot\vec{j}^{\mathrm{NC}}$,  provides the required energy input for anomalous (turbulent) heating of both electrons and ions \citep{Buchert:Effect2006}. In Sect.~\ref{General consideration} we prove this earlier conjecture using first principles with virtually no approximations. Specific fluid-model calculations for arbitrarily magnetized plasma in the quasilinear approximation show how exactly the deposited field energy is distributed between electrons and ions. This yields explicit NC expressions given by Eq.~(\ref{j_NL}), and turbulent sources of anomalous electron and ion heating given by Eq.~(\ref{diag_psi}). It has been known for a long time that anomalous electron heating (AEH) is largely a 3-D effect caused by turbulent electric fields parallel to $\vec{B}$ \citep{StMaLaher:85,Providakes:88,DimantMilikh:JGR03,
Bahcivan:Parallel2006}. According to the new development, this
requires significantly larger
$\vec{E}_{0}\cdot\vec{j}^{\mathrm{NC}}$ in 3-D than in 2-D. Due
to the mirror symmetry along $\vec{B}$, neither $\vec{E}_0$ nor
$\vec{j}^{\mathrm{NC}}$ have any significant components in the
$\vec{B}$ direction. The difference in
$\vec{E}_{0}\cdot\vec{j}^{\mathrm{NC}}$ between the 3-D and 2-D
is entirely due to noticeably larger 3-D NC caused by larger
density perturbations, as explained in the text below
Eq.~(\ref{diag_psi_summa_ei}). This prediction has been
confirmed by our recent 2-D and 3-D PIC simulations
\citep{Oppenheim:Fully2011}. As discussed in the companion
paper \citep{Dimant:Magnetosphere2011}, a strong NC, either by
itself or combined with AEH
\citep{MilikhDimant:JGR03,MilikhGoncharenko:Anom2006},  can
significantly increase the global ionospheric conductances.
This may have serious implications for predictive modeling of
MI coupling and space weather.

\section*{Appendix: Validation of Electrostatic Appoximation\label{Direct Checking}}

Using some relations obtained in Sect.~\ref{General
consideration}, we now discuss the validity of the
electrostatic approximation in describing wave processes in the
lower ionosphere. To the best of our knowledge, the
electrostatic approximation for the \emph{E}-region
instabilities has always been employed but no detailed analysis
of its validity been done, especially for the general span of
magnetization conditions considered in this paper. There are
occasions when induction fields in the ionosphere play an
important role
\citep[e.g.,][]{Amm:Towards2008,Vanhamaki:Role2007}. The main
reason why one might question the validity of electrostatic
approximation in the description of
\textit{E}/\textit{D}-region wave processes is that even tiny
induction corrections to the wave electric field along
$\vec{B}$ could modify the small parallel turbulent electric
field largely responsible for AEH.

To estimate a possible non-electrostatic component of a
separate wave field harmonic, $\delta\vec{E}_{\vec{k},\omega}$,
we split it into the electrostatic (curl-free) part,
$\delta\vec{E}_{\vec{k},\omega}^{\mathrm{ES}}\equiv(\vec
{k}\cdot\delta\vec{E}_{\vec{k},\omega})\vec{k}/k^{2}$, and the
induction part,
$\delta\vec{E}_{\vec{k},\omega}^{\mathrm{IND}}\equiv\delta\vec{E}_{\vec
{k},\omega}-\delta\vec{E}_{\vec{k},\omega}^{\mathrm{ES}}=-\vec{k}\times
(\vec{k}\times\delta\vec{E}_{\vec{k},\omega})/k^{2}$. Then
Eq.~(\ref{del_E}) yields
\begin{subequations}
\label{E_split}%
\begin{align}
\delta\vec{E}_{\vec{k},\omega}^{\mathrm{ES}}  &  =-\ \frac{i(\vec{k}%
\cdot\delta\vec{j}_{\vec{k},\omega})\vec{k}}{\varepsilon_{0}\omega k^{2}%
},\label{E_ES}\\
\delta\vec{E}_{\vec{k},\omega}^{\mathrm{IND}}  &  =-\ \frac{i\omega\vec
{k}\times(\vec{k}\times\delta\vec{j}_{\vec{k},\omega})}{\varepsilon_{0}%
(k^{2}c^{2}-\omega^{2})k^{2}}. \label{E_Ind}%
\end{align}
\end{subequations}
The wave current density, $\delta\vec{j}_{\vec{k},\omega}$, is
determined by an anisotropic response to the total turbulent
electric field, $\delta\vec {E}_{\vec{k},\omega}$. Even if the
turbulent field is largely electrostatic,
$\delta\vec{E}_{\vec{k},\omega}\simeq\delta\vec{E}_{\vec
{k},\omega}^{\mathrm{ES}}\parallel\vec{k}$, this anisotropy can
give rise to a non-negligible component of
$\delta\vec{j}_{\vec{k},\omega}$ which is not parallel to
$\vec{k}$. According to Eq.~(\ref{E_Ind}), this component
generates non-zero $\vec{k}
\times(\vec{k}\times\delta\vec{j}_{\vec{k},\omega})=\vec{k}(\vec{k}\cdot
\delta\vec{j}_{\vec{k},\omega})-k^{2}\delta\vec{j}_{\vec{k},\omega}$,
i.e., finite $\delta\vec{E}_{\vec{k},\omega}^{\mathrm{IND}}$.

Using a perturbation technique in which the zero-order wave
field is electrostatic,
$\delta\vec{E}_{\vec{k},\omega}\approx\delta\vec{E}_{\vec{k},\omega
}^{\mathrm{ES}}=-i\vec{k}\delta\Phi_{\vec{k},\omega}$, one can
easily estimate the next-order induction component,
$\delta\vec{E}_{\vec{k},\omega }^{\mathrm{IND}}$. Considering
the wave with $\vec{k}$ in the $x,z$-plane
($\hat{z}||\vec{B}_0$), neglecting density perturbations, and
applying to $\delta\vec{E}_{\vec{k},\omega}$ the regular linear
conductivity, we obtain
\[
\delta\vec{j}_{\vec{k},\omega}\approx\overleftrightarrow{\sigma^{0}}%
\cdot\delta\vec{E}_{\vec{k},\omega}^{\mathrm{ES}}=-i(\overleftrightarrow
{\sigma^{0}}\cdot\vec{k})\delta\Phi_{\vec{k},\omega}=-i\delta\Phi_{\vec
{k},\omega}\left[
\begin{array}
[c]{c}%
\sigma_{\mathrm{P}}^{0}k_{\perp}\\
-\sigma_{\mathrm{H}}^{0}k_{\perp}\\
\sigma_{\parallel}^{0}k_{\parallel}%
\end{array}
\right]  ,
\]
where we have restricted ourselves to the regular conductivity
tensor $\overleftrightarrow{\sigma^{0}}$
\citep{Kelley:Ionosphere2009},
\begin{subequations}
\label{sigma_II,P,H}%
\begin{align}
\sigma_{\parallel}^{0}  &  \equiv\frac{\vec{j}_{0\parallel}\cdot\vec
{E}_{0\parallel}}{E_{0\parallel}^{2}}=\frac{(\kappa_{e}+\kappa_{i})ne}%
{B},\label{sigma_II}\\
\sigma_{\mathrm{P}}^{0}  &  \equiv\frac{\vec{j}_{0\perp}\cdot\vec{E}_{0\perp}%
}{E_{\perp}^{2}}=\frac{(\kappa_{e}+\kappa_{i})(1+\kappa_{i}\kappa_{e}%
)ne}{(1+\kappa_{e}^{2})(1+\kappa_{i}^{2})B},\label{sigma_P}\\
\sigma_{\mathrm{H}}^{0}  &  \equiv\frac{\vec{j}_{0\perp}\cdot(\vec{E}%
_{0}\times\hat{b})}{E_{0\perp}^{2}}=-\ \frac{\left(  \kappa_{e}^{2}-\kappa
_{i}^{2}\right)  ne}{(1+\kappa_{e}^{2})(1+\kappa_{i}^{2})B}. \label{sigma_H}%
\end{align}
\end{subequations}
Since the anomalous conductivity discussed in the companion paper
is of the same order of magnitude at most, then according to
Eq.~(\ref{E_split}) we obtain
\begin{align}
\delta\vec{E}_{\vec{k},\omega}^{\mathrm{ES}}  &  \simeq-\ \frac
{(\sigma_{\mathrm{P}}^{0}+\sigma_{\parallel}^{0}k_{\parallel}^{2}/k_{\perp
}^{2})\delta\Phi_{\vec{k},\omega}}{\varepsilon_{0}\omega}\left[
\begin{array}
[c]{c}%
k_{\perp}\\
0\\
k_{\parallel}%
\end{array}
\right]  ,\label{ES_est}\\
\delta\vec{E}_{\vec{k},\omega}^{\mathrm{IND}}  &  \simeq-\ \frac{\omega
\delta\Phi_{\vec{k},\omega}}{\varepsilon_{0}k^{2}c^{2}}\left[
\begin{array}
[c]{c}%
-\ \frac{\sigma_{\parallel}^{0}k_{\parallel}^{2}}{k_{\perp}}\\
-\sigma_{\mathrm{H}}^{0}k_{\perp}\\
\sigma_{\parallel}^{0}k_{\parallel}%
\end{array}
\right]  , \label{Ind_est}%
\end{align}
where we have neglected $\sigma_{\mathrm{P}}^{0}$ compared to $\sigma
_{\parallel}^{0}$ and $k_{\parallel}^{2}$ compared to $k_{\perp}^{2}\approx
k^{2}$. We also have taken into account that in \textit{E/D}-region processes the wave phase speed, $V_{\mathrm{ph}}=\omega/k$, is many
orders of magnitude less than the speed of light, so that $\omega^{2}$ in the denominator of Eq.~(\ref{E_Ind}) can always be neglected compared to $k^{2}c^{2}$.

Comparing Eqs.~(\ref{ES_est}) and (\ref{Ind_est}) for both
parallel and perpendicular to $\vec{B}_{0}$ components and
presuming
$\sigma_{\parallel}^{0}k_{\parallel}^{2}/k_{\perp}^{2}\lesssim\sigma
_{\mathrm{P}}^{0}$, we see that the smallness of
$|\delta\vec{E}_{\vec{k},\omega}^{\mathrm{IND}}/\delta\vec{E}_{\vec{k},\omega}
^{\mathrm{ES}}|$
is ensured by
\begin{equation}
\frac{V_{\mathrm{ph}}}{c}\ll\left(  \frac{\sigma_{\mathrm{P}}^{0}}%
{\sigma_{\parallel}^{0}}\right)  ^{1/2},\qquad\frac{V_{\mathrm{ph}}}{c}%
\ll\left(  \frac{\sigma_{\mathrm{P}}^{0}}{\sigma_{\mathrm{H}}^{0}}\right)
^{1/2}. \label{provided}%
\end{equation}
The first inequality pertains to the parallel components of the turbulent
electric field, $\delta\vec{E}_{\vec{k},\omega\parallel}$, while the second,
less restrictive, inequality pertains to the corresponding perpendicular
components, $\delta\vec{E}_{\vec{k},\omega\perp}$. As mentioned above, breaking
 of the parallel-field condition may be of
importance because $\delta\vec{E}_{\vec{k},\omega\parallel}$ is crucial for
AEH. For magnetized electrons, up to the top electrojet where $\kappa_{e}\gg1$
and $\kappa_{i}\sim1$, we have $\sigma_{\mathrm{P}}^{0}\sim\kappa_{i}%
ne/B$, $\sigma_{\mathrm{H}}^{0}\sim ne/B$, $\sigma_{\parallel}^{0}%
\approx\kappa_{e}ne/B$, so that the most restrictive first
inequality becomes
\begin{equation}
\frac{V_{\mathrm{ph}}}{c}\ll\left(  \frac{\kappa_{i}}{\kappa_{e}}\right)
^{1/2}=\Theta_{0}\simeq1.4\times10^{-2}. \label{most_restrictive}%
\end{equation}
Even for an extremely strong convection electric field of
$E_{0}\simeq 150$\ mV/m, the wave phase velocity may reach
$3$~km/s at most, $V_{\mathrm{ph}}/c\simeq10^{-5}$, so that
this condition holds with a big reserve. Thus for typical
\textit{E/D}-region wave processes the electrostatic
approximation  is always valid even for small parallel
electric-field components, $\delta\vec{E}_{\vec{k},\omega||}$.

\begin{acknowledgments}
This work was supported by National Science Foundation
Ionospheric Physics Grants No.~ATM-0442075 and ATM-0819914.
\end{acknowledgments}


\end{article}
\end{document}